\documentclass[twocolumn]{article}
\usepackage{glossaries}
\usepackage{aasmacros}
\makeglossaries
\newcommand{\alf}{Alfv\'{e}n}
\newacronym[shortplural=PUIs,longplural=Pickup Ions]{pui}{PUI}{Pickup Ion}
\newcommand{\pui}{\gls{pui}}
\newcommand{\puis}{\glspl{pui}}
\newacronym[shortplural=VDFs,longplural=Velocity distribution Functions]{vdf}{VDF}{Velocity Distribution Function}
\newcommand{\vdf}{\gls{vdf}}
\newcommand{\vdfs}{\glspl{vdf}}
\newacronym[shortplural=PADs,longplural=Pitch Angle Distributions]{pad}{PAD}{Pitch Angle Distribution}
\newacronym{ism}{ISM}{Interstellar Medium}
\newacronym{lism}{LISM}{Local Interstellar Medium}
\newcommand{\lism}{\gls{lism}}
\newacronym{lic}{LIC}{Local Interstellar Cloud}
\newacronym[shortplural=ISNs,longplural=interstellar neutrals]{isn}{ISN}{interstellar neutral}
\newcommand{\isn}{\gls{isn}}
\newcommand{\isns}{\glspl{isn}}
\newacronym[shortplural=PSpDs,longplural=phase space densities]{pspd}{PSpD}{phase space density}
\newacronym{sw}{SW}{solar wind}
\newcommand{\sw}{\gls{sw}}
\newacronym{imf}{IMF}{Interplanetary Magnetic field}
\newacronym{dht}{dHT}{de-Hoffmann-Teller}

\newacronym{sws}{SWS}{Solar Wind Sector}
\newacronym{wapssd}{WAPSSD}{Wide angle partition SSD}
\newacronym{wapnossd}{WAPNOSSD}{Wide angle partition NO SSD}
\newacronym[shortplural=SSDs,longplural=solid state detectors]{ssd}{SSD}{solid state detector}
\newcommand{\ssd}{\gls{ssd}}
\newcommand{\ssds}{\glspl{ssd}}

\newacronym{pha}{PHA}{Pulse Height Analysis}
\newcommand{\pha}{\gls{pha}}
\newacronym{threed}{3d}{three dimensional}
\newcommand{\threed}{\gls{threed}}

\newacronym[shortplural=DCs,longplural=double coincidences]{dc}{DC}{double coincidence}
\newacronym[shortplural=DCRs,longplural=double coincidence rates]{dcr}{DCR}{double coincidence rate}

\newacronym[shortplural=TCs,longplural=triple coincidences]{tc}{TC}{triple coincidence}
\newacronym[shortplural=TCRs,longplural=triple coincidence rates]{tcr}{TCR}{triple coincidence rate}

\newacronym{esa}{ESA}{electrostatic analyser}

\newacronym{dsn}{DSN}{Deep Space Network}

\newacronym{dpu}{DPU}{Data Processing Unit}

\newacronym[shortplural=SIRs,longplural=stream interaction regions]{sir}{SIR}{stream interaction region}

\newacronym[shortplural=CIRs,longplural=corotating interaction regions]{cir}{CIR}{corotating interaction region}

\newacronym[shortplural=CMEs,longplural=coronal mass ejections]{cme}{CME}{coronal mass ejection}

\newacronym[shortplural=MDNVDFs,longplural=Magnetic field Direction Normalised velocity distribution functions]{mdnvdf}{MDNVDF}{Magnetic field Direction Normalised velocity distribution function}
\newcommand{\mdnvdf}{\gls{mdnvdf}}
\newcommand{\mdnvdfs}{\glspl{mdnvdf}}

\newacronym[shortplural=dMDNVDFs,longplural=differential \mdnvdfs]{dmdnvdf}{dMDNVDF}{differential \mdnvdf}

\newacronym[shortplural=QLTs,longplural=Quasi Linear Theories]{qlt}{QLT}{Quasi Linear Theory}

\newacronym[shortplural=FOVs,longplural=Fields of View]{fov}{FOV}{Field Of View}
\newcommand{\fov}{\gls{fov}}

\newacronym[shortplural=UVs,longplural=ultraviolets]{uv}{UV}{ultraviolet}
\usepackage{natbib,graphicx}
\usepackage{mhchem}
\usepackage{ulem,gensymb}

\newcommand{\sun}{\odot}
\usepackage[colorlinks=true, linkcolor=blue, citecolor=blue, filecolor=blue, urlcolor=blue]{hyperref}

\title{Three dimensional \ce{He^+} pickup ion velocity distribution functions observed with STEREO-A PLASTIC}
\author{Duncan~Keilbach$^1$
\and Verena~Heidrich-Meisner$^2$
\and Lars~Berger$^2$
\and Robert~F.~Wimmer-Schweingruber$^2$\\\small
1: TNG Stadtnetz GmbH Kiel, 2: Institut für Experimentelle und Angewandte Physik, Christian Albrechts Universität zu Kiel}
\begin{document}
\maketitle
\glsresetall

\abstract{
Freshly injected interstellar \puis are expected to exhibit a simple, torus-shaped velocity distribution function. The \pui velocity in the solar wind frame depends on the velocity of the \isn population at the pick-up position. 
}{
In this study, we aim to compare \pui velocity distributions measured by the PLasma And SupraThermal Ion Composition (PLASTIC) instrument over the full orbit of Solar TErestrial RElations Observatory-Ahead (STEREO-A) directly. 
}{
The STEREO-PLASTIC-A PUI observations are re-analysed wherein, instrumental effects of the limited \fov are accounted for. We then define a new position-independent velocity measure for \puis that takes the local direction of the interstellar neutral inflow into account. The resulting new \pui velocity measure corrects thereby for the position-dependent contribution of the \isn velocity. Each position in the orbit of STEREO can be reached by \isns following one of two trajectories, which we call primary and secondary trajectories. Therein, \isns following the primary trajectory have a higher probability to reach the location before they are ionised than particles following the secondary trajectory. Our new \pui velocity measure can be applied based on the assumption that all particles followed the primary trajectory or that all particles followed the secondary trajectory. Pitch-angle distributions are then analysed depending on the magnetic-field azimuthal angle for different orbital positions and different values of the \pui velocity measure.
}{
The new velocity measure, $\varpi_\mathrm{inj,p}$,   shows an approximately constant cut-off over the complete orbit of STEREO-A. A torus signature is visible everywhere. Therein, a broadening of the torus signature outside the focusing cone and crescent regions and for lower $\varpi_\mathrm{inj,p}$ is observed. In addition, we illustrate the symmetry between the primary and secondary \isn trajectory in the vicinity of the focusing cone. 
}{
A torus signature associated with freshly injected \puis is visible over the complete orbit of STEREO-A with increased density in the focusing cone. At least remnants of a torus signature remain for smaller values of the \pui velocity measure. The new velocity measure also prepares for \pui studies with Solar Orbiter.
}



\section{Introduction}
\label{ch:intro}
The \sw is a continuous flow of charged particles that emanate from the Sun \cite[]{Biermann1957}. The particles constitute a magnetised plasma that governs the interplanetary space surrounding the Sun. Surrounding the \sw is the \lism which itself is a magnetised plasma and therefore can not mix with the \sw. 

Interstellar \puis are a species of particles found in the \sw, which is not of solar origin, but is implanted in the \sw. They originate from \isns which inflow into the heliosphere. The trajectory of an interstellar neutral is mainly governed by two forces: gravitation from the Sun \cite[]{Axford1972,VS1976} (and to a for the purpose of this paper negligible degree from the larger planets of the solar system) and radiation pressure from solar ultraviolet photons (UVs) \cite[]{Tarnopolski2009,Shestakova2015}. Radiation pressure is relevant for, for example, protons, however almost negligible for \ce{He} particles. \ce{He} exhibit at 1 AU (astronomical unit) a prominent focusing cone behind the Sun \citep{Gloeckler2004} and a crescent feature on the other side \citep{McComas2004,drews2012inflow,Sokol2016}.

Without the influence of radiation pressure, the trajectories of \isns are solutions to the two-body problem and therefore elliptical trajectories \cite[]{Axford1972}.

Since the computation of these trajectories has been done before by several studies, this study focuses on the consequences of the \isn velocities before pick up on the \pui velocity characteristics. This study's method to compute the trajectories and velocities is found in Appendix \ref{ch:isnmod}. A consequence of the symmetry of the two-body problem is the formation of the focusing cone. Here, enhanced \isns densities are expected \cite[]{Axford1972}. Yet, neutral particles can not be measured by a majority of space-borne instruments which are built for the detection of charged particles. Still, the measurement of interstellar neutral particles or neutral particles in the heliosphere in general has led to the discovery of the IBEX ribbon (Interstellar Boundary Explorer, \citet{McComas2009}). %
Indirectly, through \puis, \isns can still be detected: On its trajectory through the heliosphere an \isn is likely to become ionised, most prominently for He by photoionization by ultraviolet radiation. Therefore, the ionization probability increases with decreasing distance to the Sun. Through ionization, the former neutrals are subjected to the local magnetic field's Lorentz Force which causes them to gyrate around the magnetic field lines. These are embedded into the outward propagating \sw. Hence, the particles are picked up by the SW and called \puis.

Compared to ions heavier than \ce{H} or \ce{He} in the \sw from the Sun, most \puis undergo only a single or, more seldom a double ionization \cite[]{Axford1972,Gloeckler1993,Gloeckler1995h,Moebius1985, Gloeckler1998,Gloeckler2000}. Therefore, \puis can in many cases be distinguished from the \sw by their charge state. This study focuses on \ce{He+} \puis. Since most \ce{He} of solar origin is fully ionised in the \sw, the charge state is already a good indicator for a \pui. However, for \ce{He^{2+}} \puis, characteristics of the \vdf are the only information to distinguish their in-situ observations from ions of the \sw. For \ce{He+} \puis, compared to different ion species, their distinct mass-per-charge ratio of 4 is also well distinguished from different ion species which makes them an ideal candidate for studies with time-of-flight-mass-spectrometers (Sect.~\ref{ch:heplusprep}). 

\begin{figure*}
\includegraphics[width = \textwidth]{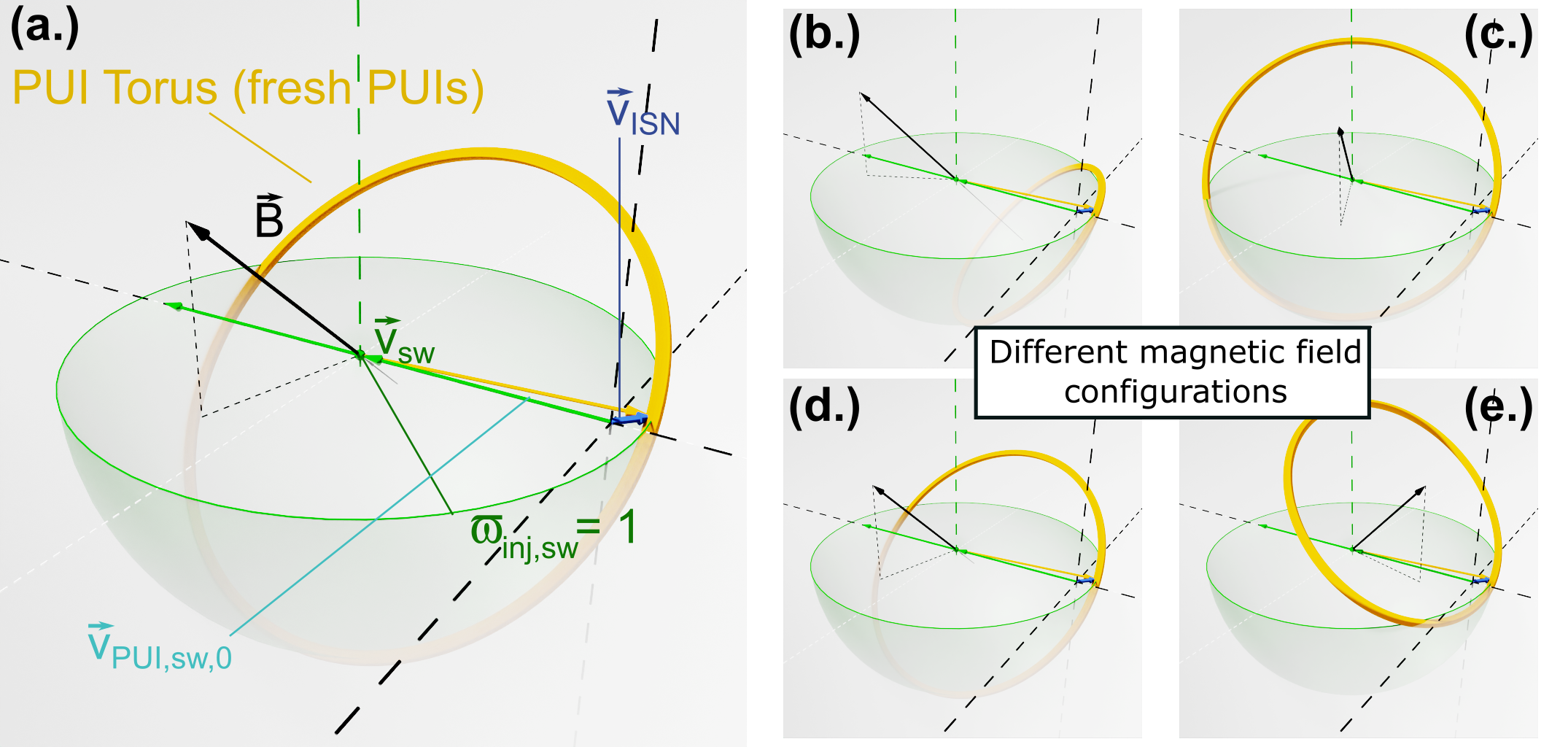}
\caption{Three dimensional visualization of a freshly injected \pui in velocity space in the SW frame of reference. Green arrows represent the \sw bulk velocity, black arrows the local magnetic field, blue arrows the \isn velocity and the amber ring the initial \pui \vdf. The green hemisphere is a cut out of the sphere of the entirety of velocities of fresh \puis at the depicted setup of \sw velocity and \isn velocity. The different visualizations depict in total five different magnetic field configurations while the other parameters are kept constant.}
\label{fig:PUIinjsketch}
\end{figure*}

Fig.~\ref{fig:PUIinjsketch} depicts the injection of a \pui in \threed velocity space for different magnetic field configurations. The initial \isn has a velocity of $\vec{v}_\mathrm{ISN}$ on its Keplerian trajectory. However, in the \sw frame (green arrow) that transports the magnetic field, the initial velocity of a newly created \pui is the difference between the \isn velocity and the frame's velocity $\vec{v}_\mathrm{SW}$. It is important to notice, that the $\vec{v}_\mathrm{SW}$ is not necessarily the correct velocity of the frame of transport of charged particles in the \sw. It might be shifted by a velocity of the magnitude of the local \alf velocity \citep{Nemecek2020}. However, a detailed analysis of the transport velocity is beyond the scope of this work and might even require better precision than can be provided with the employed instruments. In the transporting frame, the vector $\vec{v}_\mathrm{SW} - \vec{v}_\mathrm{ISN}$ is the origin of the gyration around the local magnetic field (the difference between the blue interstellar neutral speed and the green solar wind speed in Fig~\ref{fig:PUIinjsketch}). This results in a circular trajectory in velocity space with a radius of $v_\perp$ with $v_\perp = \hat{e}_\mathrm{B} \cdot (\vec{v}_\mathrm{SW} - \vec{v}_\mathrm{ISN})$, where $\hat{e}_\mathrm{B}$ is a unit vector parallel to the magnetic field (amber ring in Fig.~\ref{fig:PUIinjsketch}). In addition to the gyration, the \pui may move parallel to the magnetic field with a velocity of $v_\parallel$. The pitch angle, that is the inclination angle of the particle's gyration trajectory with respect to the local magnetic field is a function of the angle between the local magnetic field and the initial \pui velocity and can be expressed as $\alpha = \arctan(v_\perp / v_\parallel)$. Naturally, each circle centred around the local magnetic field in velocity space is a collection of velocities of the same pitch angle. Under the assumption that multiple \puis are injected with the same or similar pitch angles at slightly different times, a collection of gyrating particles at different phases is found. Therefore for freshly created \puis, a torus distribution is expected in velocity space.

Yet, it is often assumed, that during transport in the \sw, the pitch angle distribution becomes less anisotropic quickly \cite[]{VS1976,Isenberg1997}. Previous studies with one-dimensional velocity data observed anisotropies \cite[]{Moebius1995,Gloeckler1998}. In contrast, clear signs of remains of torus-shaped distributions were observed and matched with the magnetic field orientation in two-dimensional velocity data with the PLasma And SupraThermal Ion Composition (PLASTIC) instrument \citep{Drews2015} aboard the Solar TErestrial RElations Observatory-Ahead (STEREO-A)  and with the Magnetospheric MultiScale (MMS) \citep{2021ApJ...913..112S}. For a more detailed analysis, this study aims to employ \threed velocity information. The mentioned studies also emphasise that there is an important factor to understanding the relationship between \puis and the \isns they originate from: The strength of transport effects in the \sw or as a proxy the time scale of isotropisation in the \sw.

A method to classify the time scale of transport modifications, is to compare freshly created \puis with \puis which had been injected in the \sw earlier. After pick-up, the newly generated \puis gyrate around the magnetic field direction. The interplanetary magnetic field itself moves with the solar wind velocity. Therefore, if we, for the moment, neglect the initial \isn velocity, the relative velocity of the newly generated \pui to the solar wind frame is also the solar wind velocity \citep{Gloeckler1998,Gloeckler2004,Drews2015}. Since the initial velocity of \puis is thereby known, the velocity $v_\mathrm{PUI}$ can be employed as a qualifier for the recentness of ionization of a \pui. Hence, the quantity $\varpi_\mathrm{SW}$ with
\begin{equation}
\label{eq:wrefswdef}
\varpi_\mathrm{SW} = \frac{|\vec{v}_\mathrm{PUI} - \vec{v}_\mathrm{SW}|}{|\vec{v}_\mathrm{SW}|} 
\end{equation}
has been introduced for the analysis of \pui velocities \cite[]{Gloeckler1998}. Neglecting the \isn velocity, a fresh \pui is expected with a $\varpi_\mathrm{SW} = 1$. If the \isn velocity is not neglected, fresh \puis deviate from $\varpi_\mathrm{SW} = 1$, wherein the difference (or shift) is a function of the position-dependent \isn velocity and therefore a function of orbital position. \cite{Moebius2015,Taut2018} employed this shift in \pui velocities to determine the inflow direction of the \lism. It is necessary to point out, that \cite{Moebius2015} measured the position-dependent shift in \pui velocities from $\varpi_\mathrm{SW}$ spectra. Hence, the resulting shift is intrinsically a result of the directional difference between $\vec{v}_\mathrm{ISN}$ and $\vec{v}_\mathrm{SW}$. In contrast, this study proposes a method to account for the velocity shift directly with the 3d velocity information instead of with absolute velocities. To that end, Sect.~\ref{ch:winjdef} presents a new \pui velocity measure, $\varpi_\mathrm{SW},$ which includes the local \isn velocity. The clear advantage of a velocity measure that accounts for the \isn velocity is that the \vdfs at different orbital positions are disentangled from the position-dependent velocity shift and can therefore be more readily compared than without the inclusion of \isn velocity. Also, for any point in the heliosphere at least two \isn trajectories intersect under the consideration of orbital mechanics with this point (compare Appendix \ref{ch:isnmod}). Hence, their influence on \pui signatures can be investigated. Table \ref{tab:wevo} provides an overview on different \pui velocity measures and their properties.

It is again important to point out, that in contrast to, for example, \cite{Drews2015} this study aims to employ full \threed vector information for both $\vec{v}_\mathrm{PUI}$ and $\vec{v}_\mathrm{SW}$ in an effort to increase the analysis' precision and mitigate possible projection errors. Sect.~\ref{sec:torus} revisits the results from \cite{Drews2015} based on the new velocity measure and Sect.~\ref{ch:padresults} utilises the new velocity measure to compare \pui pitch angle distribution along the full orbit of STEREO-A.

\begin{figure*}
\includegraphics[width = \textwidth]{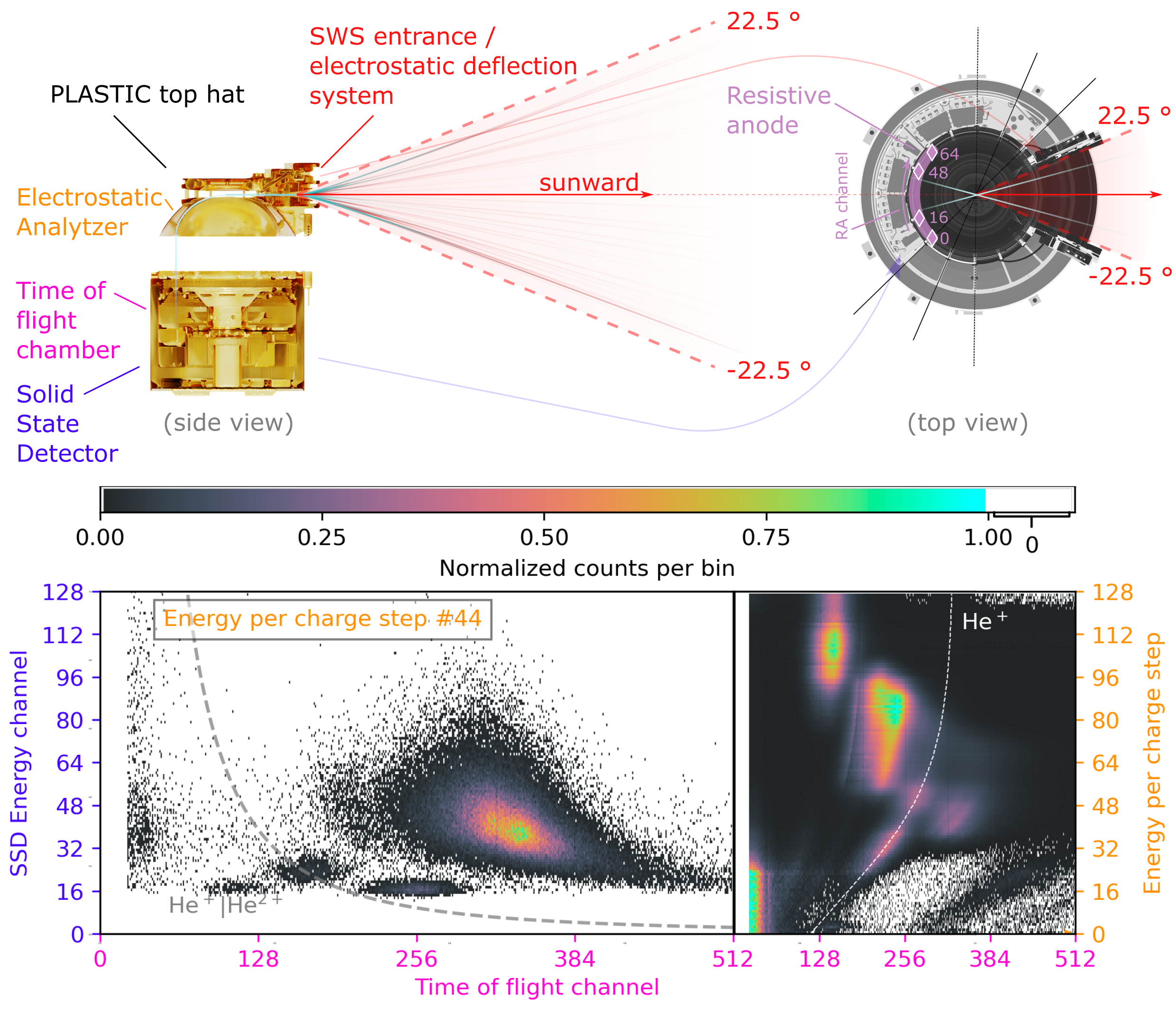}
\caption{Top left: Rendered side view of PLASTIC (cut through in half). The entrance of top-hat of the cylindrical sensor is separated into two parts: The solar wind sector and the wide angle partition. Within the top-hat, the electrostatic analyser is housed. In the lower parts of the instrument, the time-of-flight chamber and an \ssd are found. An example trajectory that enters the top-hat is plotted in turquoise. Top right: Superimposed top-view cross sections of PLASTIC. On the top, the entrance system is visible. At lower layers, the resistive anode and solid state detectors are found. In purple, resistive anode channels corresponding to the incident directions from the SWS aperture on the opposite side of the instrument are shown.
Bottom: Histograms of PLASTIC pulse height analysis data from 2009. The left histogram shows the \ssd energy channel as a function of the time-of-flight channel, wherein only events contribute which occur during energy-per-charge step 44. The gray dashed line shows the  expected positions of \ce{He+} and \ce{He^{2+}} for all energy-per-charge steps. The right histogram shows energy-per-charge step as a function of time-of-flight channel. Here, the white dashed line shows the expected average position of \ce{He+}. The right histogram is normalised to individual maxima of time-of-flight slices.}
\label{fig:PLASTICexpl}
\end{figure*}
\section{Preparation of STEREO-A PLASTIC \pha data}
For a measurement of full \pui \vdfs and an assessment of their qualitative shape, specific requirements need to be met by the measuring instrument. Firstly, the instrument needs to determine the mass and charge state of each particle. Secondly, it is necessary that the instrument is capable of measuring particles with significantly lower densities than the proton part of the \sw. Thirdly, it is necessary to not only measure 1d absolute velocities but to gain full \threed vector information of the velocities. The combination of these criteria is met by the PLasma And SupraThermal Ion Composition (PLASTIC, \citet{Galvin2008}) instrument aboard the Solar TErestrial RElations Observatory-Ahead (STEREO-A).

\begin{figure}
\includegraphics[width = \columnwidth]{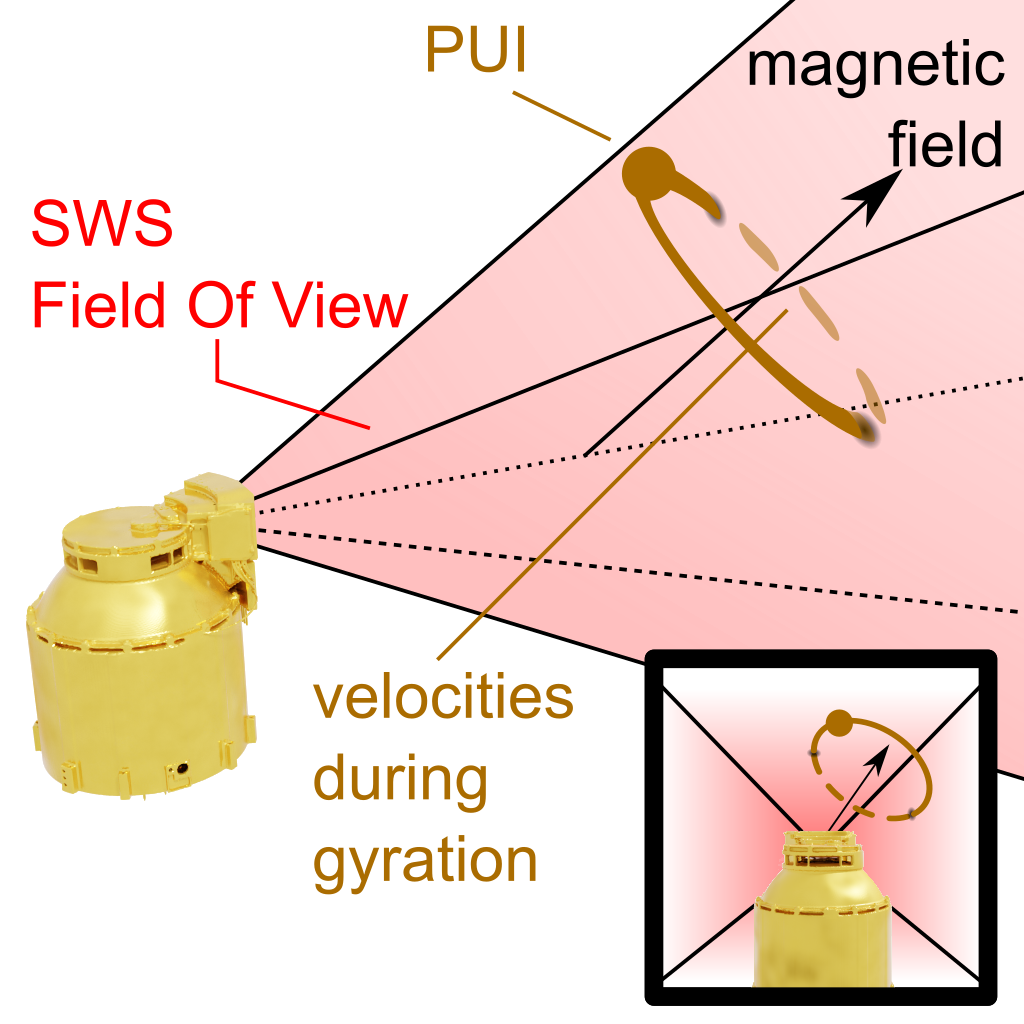} 
\caption{Artistic rendition of the PLASTIC instrument (golden), its field-of-view projected into \threed velocity space (red) and the trajectory of a \pui (turquoise) which gyrates around the magnetic field (black arrow). Parts of the \puis example trajectory are outside the \fov. So, the probability of the particle to be measured is less than 1. The inset at the bottom right provides the perspective which looks from behind PLASTIC into its \fov.}
\label{fig:ringcovap}
\end{figure}

\begin{figure*}
\includegraphics[width = \textwidth]{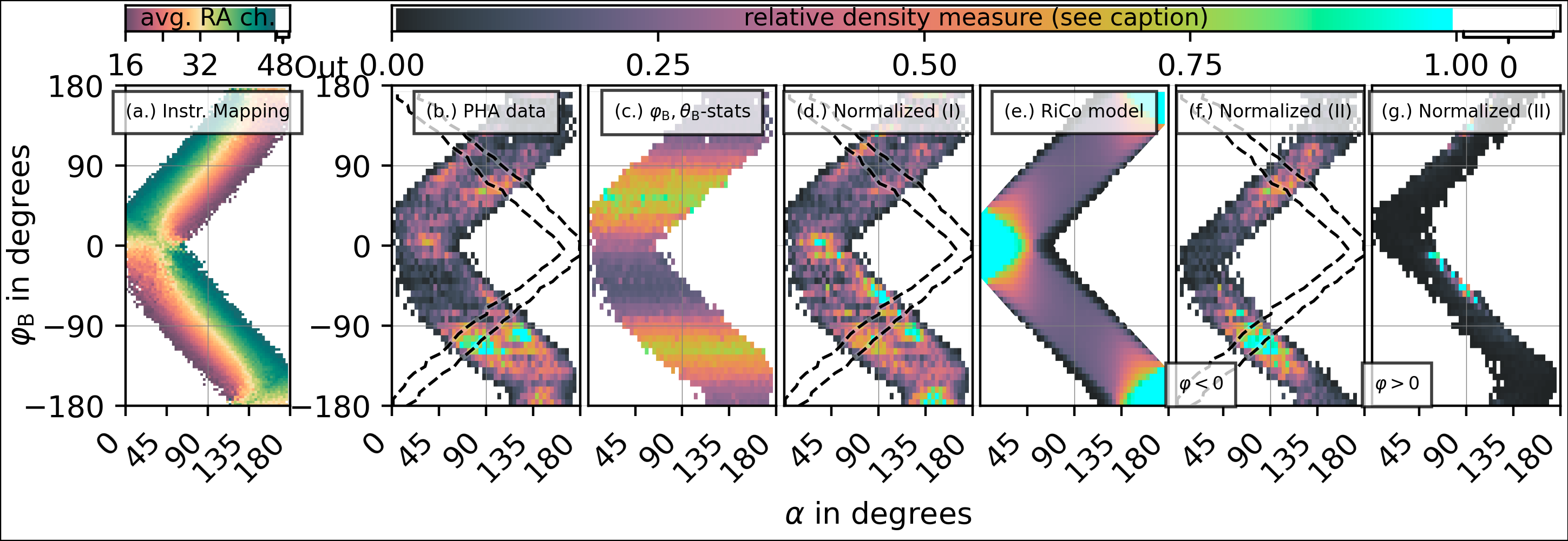}
\caption{Overview of instrumental resistive anode mapping (a.) and normalizations ((b.) to (g.)) applied to the \ce{He+} data: All plots depict a quantity as a function of pitch angle ($x$-axis) and magnetic field azimuthal angle ($y$-axis). All identified \ce{He+} events are selected, for which the conditions $0.9 < \varpi_\mathrm{inj,p} < 1.1$, $\pi/16 < \vartheta_\mathrm{B} < \pi/16$ and $|\lambda_\mathrm{ecl} - 75| < 32$ apply. Panel (a.) depicts the average position channel  measured by PLASTIC's resistive anode (average resistive anode channel, av. RA ch.). Panel (b.) depicts He$^+$ \pha data as counts per bin. Panel (c.) depicts weights based on the number of instrumental weighting cycles during which the magnetic field angle was observed. These are applied in a first normalization step. The result is shown in Panel (d.) as normalised counts per bin. Panel (e.) depicts weights based on the coverage of a particle's gyro orbit within the aperture of PLASTIC. Panels (f.) and (g.) depict the result of normalization by these weights. The data from the $\varphi<0^\circ$ and $\varphi>0^\circ$ halves of SWS are separately depicted in Panel (f.) and Panel (g.), respectively.}
\label{fig:ringcovdata}
\end{figure*}

\subsection{He$^+$ \pha data}
\label{ch:heplusprep}

Fig.~\ref{fig:PLASTICexpl} shows an overview of the structure of PLASTIC (top half) together with an example of typical observations for a single energy-per-charge step (bottom half of Fig.~\ref{fig:PLASTICexpl}).
PLASTIC is a time-of-flight mass spectrometer and a cylindrical top-hat sensor (see cross sections in Fig. \ref{fig:PLASTICexpl}) which originally was designed to measure protons and heavy ions with a lateral \fov of almost $360^\circ$. The instrument is segmented into three sections which were built for different requirements: The Solar Wind Sector (SWS) observes in the direction of the incoming \sw and not only features a triple coincidence measurement (electrostatic analyser, time-of-flight, and \ssd) but the capability of elevation angle measurements with an electrostatic deflection system. Herein, a range of $45^\circ$ is separated into 32 bins. In the SWS, lateral direction resolution is achieved with a resistive anode with also 32 bins (channels 16-48) in a range of $45^\circ$ within the SWS. The other two sections were built with a coarser binning in the longitudinal incident angle, do not feature an elevation angle measurement, and are partly not equipped with \ssds. Hence, in this work only SWS data is employed.

The pitch angles covered by the position channels of the resistive anode are depicted in Fig. \ref{fig:ringcovdata}(a.) (which summarises all pre-processing steps and is discussed in more detail in Sect.~\ref{sec:rico}) as a function of magnetic field azimuthal angle. Unfortunately, the instrument measures particles with an asymmetric efficiency \cite[]{Keilbach2023phd}. Hence, in this study data from the $\varphi<0^\circ$ half of SWS (channels between 16 and 32) and the $\varphi>0^\circ$ half of SWS (channels between 32 and 48) are considered separately as if they would originate from different sensors and we mainly focus on the  $\varphi<0^\circ$ half. The consequence is that the pitch angle coverage as a function of magnetic field orientation becomes smaller for each instrument half  as seen from the mapping of position channels to pitch angles in Fig. \ref{fig:ringcovdata}(a.).

Also, data from position channels in proximity to channel 32 is to be viewed with additional care because particles are scattered at a support structure in the centre of the instrument. Hence, channel 32 is practically a blind spot of the instrument (the events which occur there are likely due to scattering in the instrument) and channels next to it are likely affected by edge effects.

The instrument steps through 128 energy-per-charge settings every minute. Per energy-per-charge step, the electrostatic deflection system sweeps once through its 32 settings. Since at the energy-per-charge steps corresponding to protons, the flux through the instrument increases drastically, the SWS's entrance system is split into a larger and a smaller opening (main channel and small channel). If a flux threshold is reached, the instrument switches between these two channels. In this study, the ions of interest are expected in the main channel. Further, \pha events are pre-selected by an on-board logic with priorities defined by ground-calibrated priority classes. The idea behind this is to prevent that under the limited telemetry of the instrument, the most common particle species in the \sw take most of the available telemetry to the detriment of the less common heavier ions. For the purpose of correcting the number of events, the instrument accumulates statistics of the total number of events per priority class in cycles of 5 min. Hence, instrumental data is available with a cadence of 5 min. 

The identification of \ce{He+} particles from the \pha data is described in detail in \cite{Keilbach2023phd}. The peaks of both \ce{He+} and \ce{He^{2+}} in \ssd energy vs. time-of-flight histograms (as an example the histogram for energy-per-charge step 44 is shown in Fig. \ref{fig:PLASTICexpl} bottom left) are tracked while the contributing data is filtered by energy-per-charge step. As an overview, a histogram of energy-per-charge vs. time-of-flight is depicted in the bottom right of Fig. \ref{fig:PLASTICexpl}. This leads to a model function for the ions' positions which is then employed to separate the \ce{He+} ions from the remaining \pha data. Once the ions and therefore their mass-per-charge $m/q$ are identified, the absolute velocity $v$ is obtained from the energy-per-charge $\text{\ae} = (m/q)\cdot v^2/2$. A vector quantity is gained from the absolute velocity from the information of the electrostatic deflection step and the resistive anode. The selection and calibration approach applied in this study is explained in more detail in \cite{Keilbach2023phd}. In 2015, STEREO-A was rotated by $180^\circ$ around the sun-pointing axis. We focus on the time period prior to this flip of STEREO-A, that is, we analyse data from 2008-2015. Interplanetary coronal mass ejections (ICMEs) and stream interaction regions (SIRs) have been removed from the data set based on the respective available \citep{Jian2013,Jian2018} and \citep{Jian2019} lists.

To reduce contamination of protons and \ce{O^{6+}} events, additional filtering is required. The mass-per-charge of \ce{He+} is $4$ amu/$e$, the mass-per-charge of \ce{O^{6+}} is $~2.7$ amu/$e$ and the mass-per-charge of protons is $1$ amu/$e$. So, for the contaminating elements, the mass-per-charge is considerably lower than for \ce{He+}. So, if a proton or \ce{O^{6+}} is misidentified as \ce{He+}, its velocity computed from energy-per-charge is considerably slower than that of the \sw bulk, since unfittingly the mass-per-charge of $4$ amu/$e$ is assumed for the velocity computation. For interstellar \puis, velocities in the spacecraft frame are expected which range to at least twice the \sw velocity. This leads to the criterion that all particles with $w_\mathrm{SW}:= |\vec{v}_\mathrm{PUI} - \vec{v}_\mathrm{SW}| / |\vec{v}_\mathrm{SW}| < 0.5$ and $\vec{v}_\mathrm{PUI} - \vec{v}_\mathrm{SW} / |\vec{v}_\mathrm{SW}| < 0$ are labelled as likely contamination and are disregarded. In addition, instrumental deficiencies and asymmetries have been identified. Their treatment is also described in \cite{Keilbach2023phd}.

\subsection{Combination with STEREO-A IMPACT MAG Magnetic field data}
The magnetometer (MAG) of STEREO-A is part of the In-situ Measurements of Particles and CME Transients (IMPACT) suite and is a triaxial flux gate magnetometer with a nominal cadence of $8$ Hz at a resolution of $\pm 0.1$ nT  in normal mode \cite[]{Acuna2008,Luhmann2008}. To match the cadence of PLASTIC, the vector $8$ Hz was processed and averaged to the 5-min time resolution of PLASTIC. This is further described and discussed in Appendix \ref{ch:magavg}.

In this work, \pui pitch angle distributions are investigated as a function of magnetic field angles. This requires that event distributions are filtered for different magnetic field angles. However, different magnetic field angles have been observed at STEREO-A with different frequencies and therefore, \ce{He+} \puis under different magnetic field configurations do not have the same detection probability. To provide a consistent and comparable representation, the resulting distributions are convolved with the total numbers of occurrences of the magnetic field angular configuration they were measured at. Thereby, the resulting \pui distributions are divided by the total number of $5$ min intervals during which a certain angular configuration has been measured. The total histogram of normalization factors is depicted in Fig. \ref{fig:globalmagangles}.

\subsection{Correction of pitch angle dependent measurement probability and consideration of the \fov}
\label{sec:rico}

Fig. \ref{fig:ringcovap} is an artistic rendition of PLASTIC and a projection of its \fov (red cone). The radial (sunward) direction (black line) is in the centre of the \fov. Apart from the rendition of PLASTIC itself, the sketch depicts velocity space. Hence, at a distance along the radial that represents the \sw velocity, an example magnetic field vector (black arrow) is shown and a ring corresponding to a \pui which gyrates around the field. At this point the ring should not be interpreted as an anisotropic \vdf but as the trajectory of a single particle within an arbitrary gyrotropic distribution. For this example configuration of the trajectory of a \pui, parts of the trajectory are outside the \fov and parts inside. Hence, the \pui may not be measured, if it is outside the \fov while it is close enough for measurement. So, under the assumption that for any \pui the phase of its gyration is a statistical quantity, the coverage of the gyration trajectory with the \fov is a determining factor in the measurement probability.

Thus, pitch angle distributions need to be corrected for the influence of this effect, which we call ring coverage. In the following, the ring coverage based measurement probabilities are computed and pitch angle distributions shown in this work are corrected through division by the ring coverage.

The ring coverage based measurement probabilities are computed per \pha event with a method described in detail in \cite{Keilbach2023phd}. The first step is to construct a ring which represents the trajectory and corresponds to the \pha event based on the current \sw velocity, $\varpi_{\mathrm{SW}}$, pitch angle, and the directions of the \sw velocity and magnetic field vectors. Then, the intersections of the ring with the \fov boundaries are found (if existing). If no intersections exist, the ring is either fully inside or outside of the \fov which leads to either full coverage ($1$) or no coverage ($0$). If boundaries exist, their angular distance is computed along the path that is inside the \fov. This distance divided by $2\pi$ is the ring coverage based measurement probability. The unlikely case that exactly one intersection with the \fov boundaries exists (the ring grazes the boundary) is treated as either full coverage or no coverage, respectively. In the case of more than two intersections with the boundaries, the corresponding points on the ring are separated into pairs, so that for each of these the previous deliberations for two intersections apply. The sum of the resulting coverages of those pairs represents all points. Hence, the computational pattern for more than two intersections does not need extra description. 

Fig.~\ref{fig:ringcovdata} provides an illustration of the normalization steps applied in this study. Panel (b.) in Fig.~\ref{fig:ringcovdata} depicts an example of a raw PHA derived pitch angle distribution as a function of magnetic field azimuthal angle. Panel (c.) shows the respective normalization factors based on the frequency with which each magnetic field configuration was observed and Panel (d.) gives the histogram after this normalization was applied. In Panel (e.) the average ring coverage of the contributing particles is shown. In Panels (f.) and (g.) the result of the final normalization is given for each half of the instrument: the influence of the events which contribute to Panel (d.) are weighted with the reciprocal of the ring coverage based measurement probability.

In Sect.~\ref{ch:padresults} we put the observations of this study into the context of \citet{Drews2015}. To this end, we here provide a short overview on the differences in the respective data processing. (1) \citet{Drews2015} restricted the analysis to a 2d representation of the velocity space, whereas here, we consider the full 3d case. Since (2) \citet{Drews2015} restricted its analysis to cases where the ring coverage based measurement probability is expected to be constant, such a correction was not applied there. (3) Unlike in our present study, the orbit-depending effect of the \isn velocity on the \pui velocity measure was not considered in \citet{Drews2015}. 
 
\begin{table*}
\centering
\begin{tabular}{|p{5cm}|p{5cm}|p{5cm}|}
\hline
\hline
$w_\mathrm{SW} = \frac{|\vec{v}_\mathrm{PUI}|}{|\vec{v}_\mathrm{SW}|}$ & $\varpi_\mathrm{SW} = \frac{|\vec{v}_\mathrm{PUI} - \vec{v}_\mathrm{SW}|}{|\vec{v}_\mathrm{SW}|}$ & $\varpi_\mathrm{SW,inj} = \frac{|\vec{v}_\mathrm{PUI} - \vec{v}_\mathrm{SW}|}{|\vec{v}_\mathrm{SW} - \vec{v}_\mathrm{ISN}|}$ \\
\hline
&&\\
Fresh \puis are injected to $w_\mathrm{SW} \simeq 2$.&Fresh \puis are injected to $\varpi_\mathrm{SW} \simeq 1$.&Fresh \puis are injected to $\varpi_\mathrm{SW,inj} = 1$.\\
\hline
Corresponds to concentric shells in absolute energy space.&Corresponds to \pui energy with respect to the frame of transport.&Corresponds to the tentative pickup point in velocity space.\\
&&\\
Sensitive to orbital position-dependent shift of velocity spectra.&Sensitive to orbital position-dependent shift of velocity spectra.&Accounts for $\vec{v}_\mathrm{ISN}$ and is therefore not affected by velocity shift.\\
\hline
\hline
\end{tabular}
\caption{Overview on \pui relative velocity measures.}
\label{tab:wevo}
\end{table*}

\begin{figure}[tp!]
    \centering
    \includegraphics[width=\columnwidth]{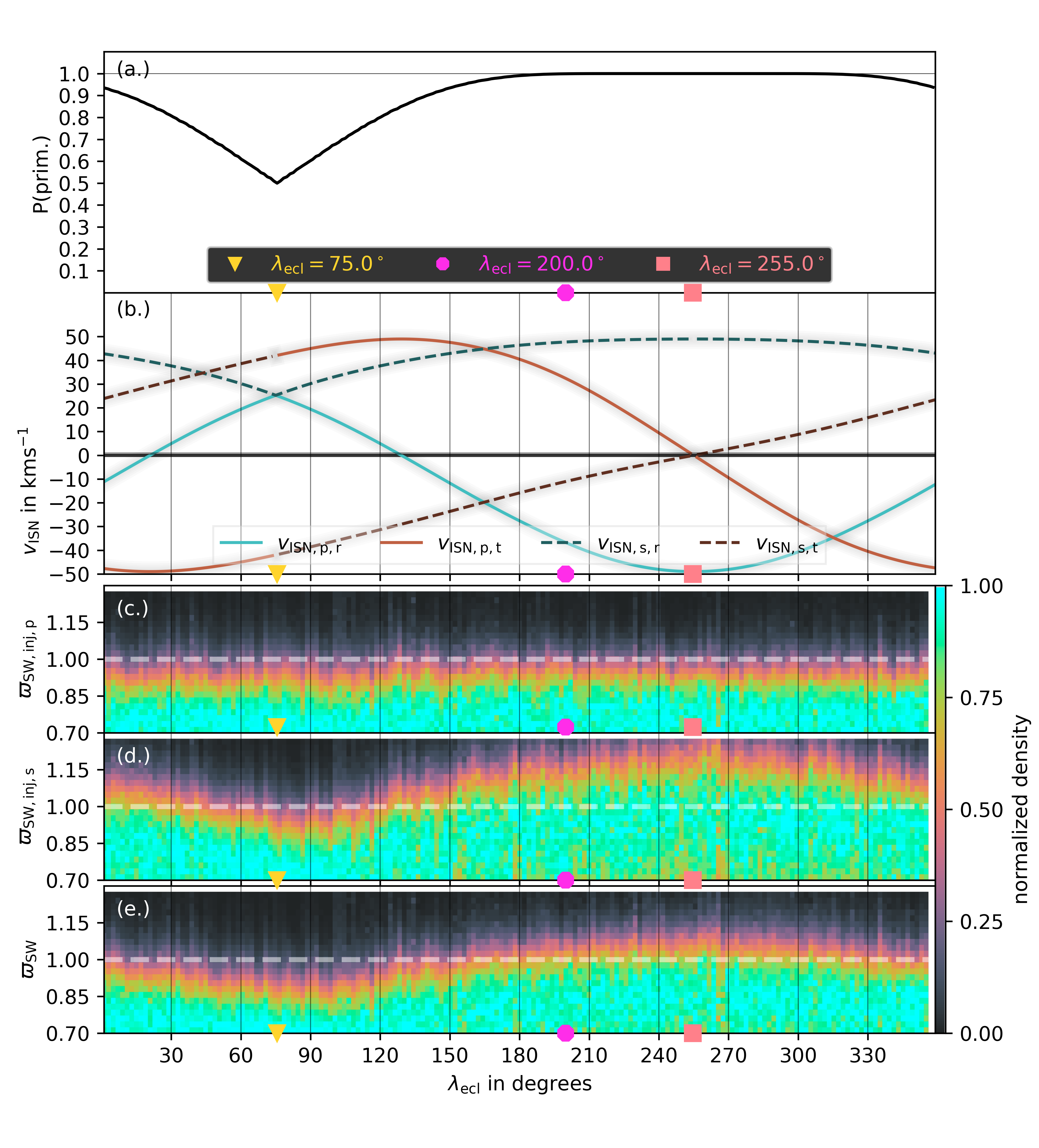}
    \caption{\isn densities and velocities and the impact of the \isn velocities on relative velocity measures for \puis over ecliptic longitude. All panels share a common $x$-axis and are a function of the ecliptic longitude. Panel (a.) shows the fraction of primary \isn density and Panel (b.) the radial and tangential components of the \isn velocity. Panel (c.) shows spectra of $\varpi_\mathrm{SW,inj,p}$, Panel (d.) of $\varpi_\mathrm{SW,inj,s}$ and Panel (e.) of $\varpi_\mathrm{SW}$. All spectra are normalised to the respective maximum of ecliptic longitude slices. The different symbols at the $x$-axes correspond to different ecliptic longitudes which correspond in heliospheric position to the markers in Fig. \ref{fig:vISNmodel}: focusing cone (yellow triangle), crescent (light pink square), and intermediate (magenta circle) position.}
    \label{fig:vISNmodelshowcase}
\end{figure}


\begin{figure}
\centering
\includegraphics[width = 0.5\textwidth]{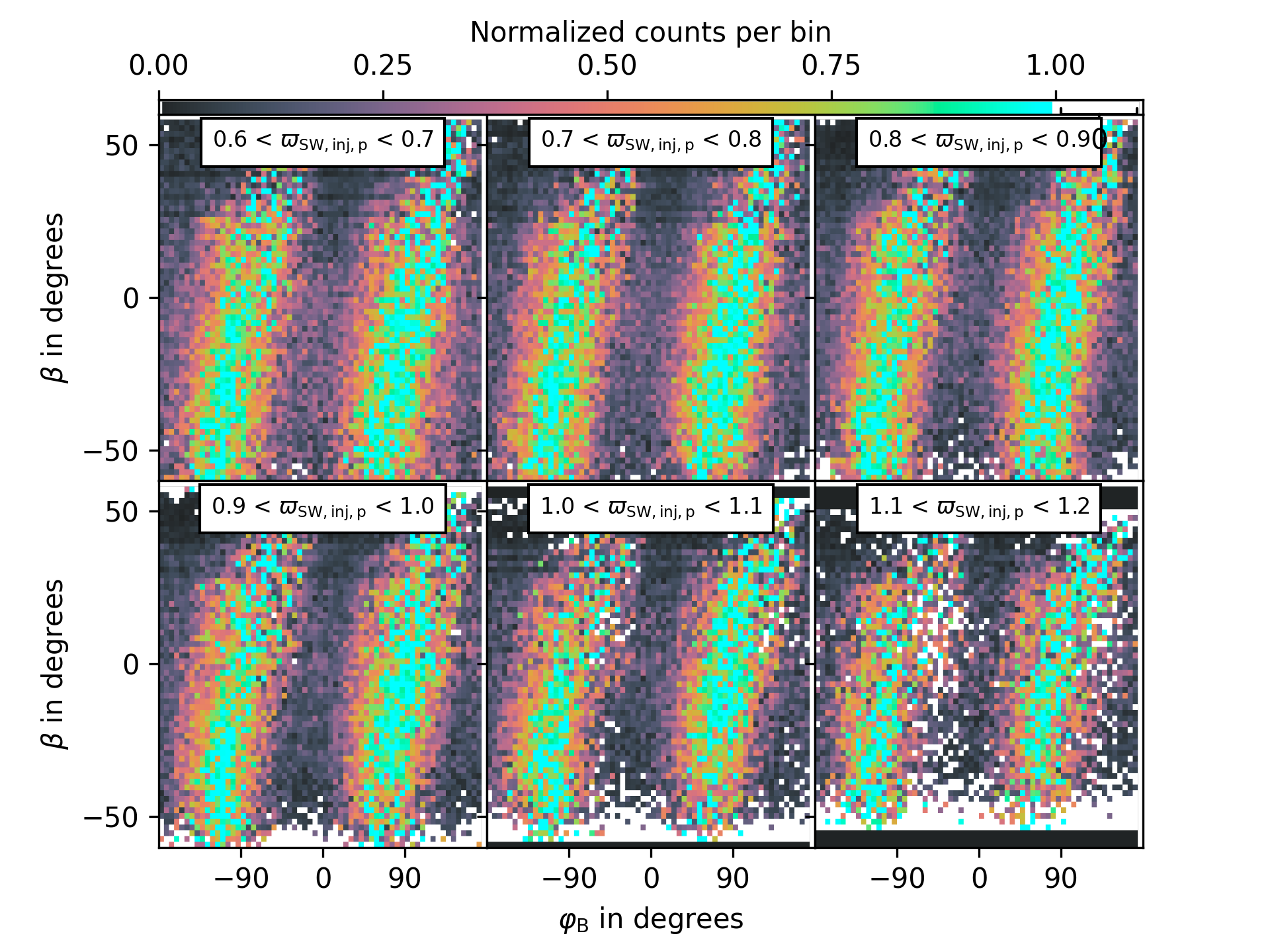}
\caption{2d histograms of the angle $\beta$ of Equation \ref{eq:beta} ($y$-axis) as a function of the magnetic field azimuthal angle ($x$-axis). Each histogram is accumulated from events from different intervals of $\varpi_\mathrm{SW,inj,p}$. Each $\beta$-slice is normalised to its respective 99-percentile.} 
\label{fig:drewstorus}
\end{figure}

\begin{figure*}[tp!]
\centering
\includegraphics[width = \textwidth]{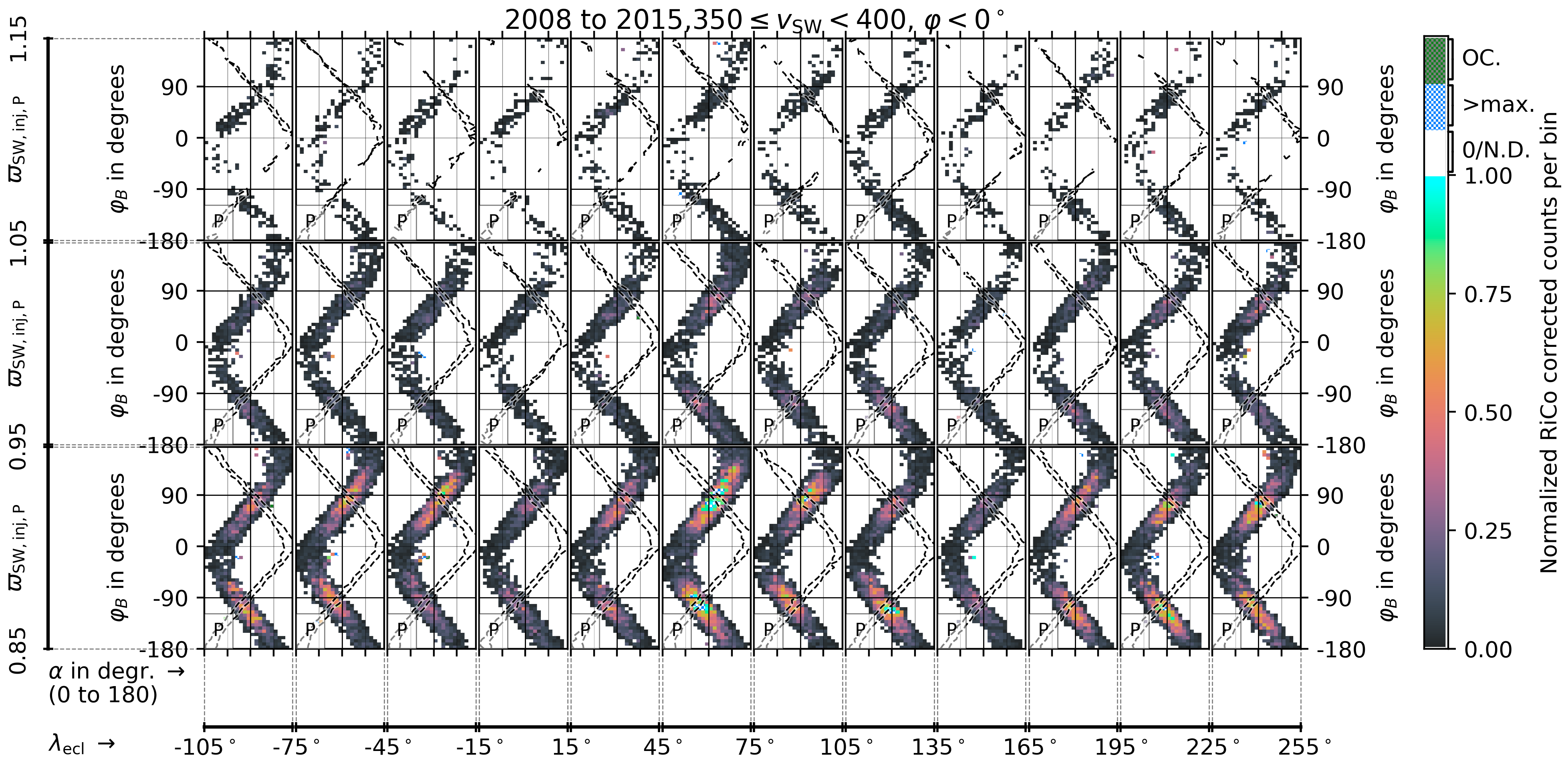}
\caption{2d histograms of pitch angle ($x-axis$, from $0^\circ$ to $180^\circ$ with one tick equalling $45^\circ$) and magnetic field azimuthal angle ($y-axis$, from $-180^\circ$ to $180^\circ$). The entirety of histograms is structured as a matrix, in each cell pitch angle histograms for the primary \isn trajectory (P) and each row of histograms is filtered for a different $\varpi_\mathrm{inj,P}$, respectively. The value range for a histogram is indicated on the left line. Each histogram is labelled P to indicate that all observations are treated as if coming from the primary trajectory. Each column of histograms represents a different ecliptic longitude which is indicated on the bottom line. All histograms share the same global maximum after the normalization steps displayed in Fig.~\ref{fig:ringcovdata} were applied. The dashed lines show the range of initial pitch angles to be expected at a given magnetic field azimuthal angle. In addition, a solar wind filter is applied ($300 \mathrm{km/s} \le v_{\mathrm{SW}} \le 450 \mathrm{km/s}$) and only data from the $\varphi<0^\circ$-half of the instrument is shown here.}
\label{fig:whistOverviewRightGlobal}
\end{figure*}

\begin{figure*}[tp!]
\centering
\includegraphics[width = \textwidth]{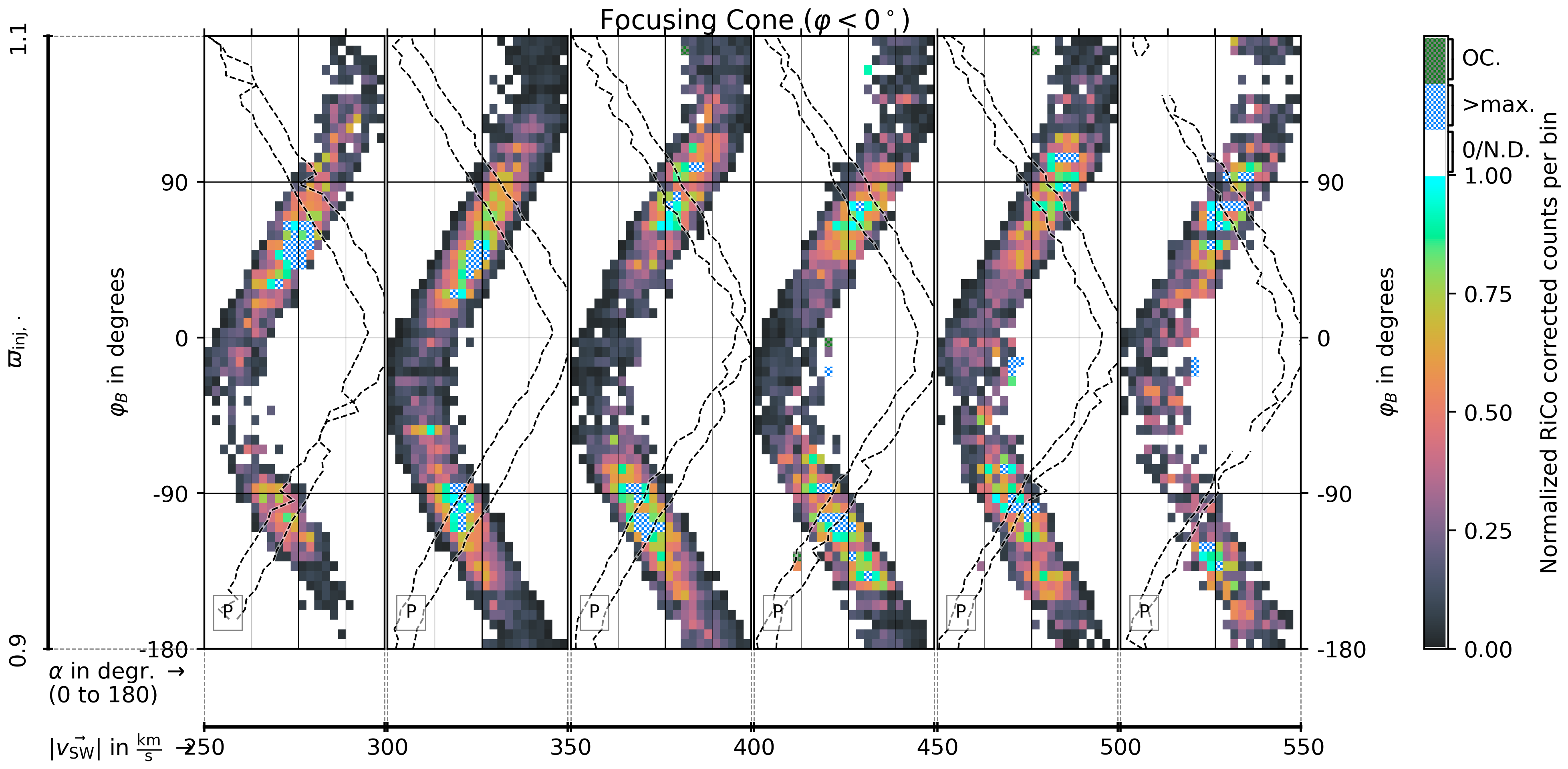}
\caption{2d histograms of pitch angle ($x-axis$, from $0^\circ$ to $180^\circ$ with one tick equalling $45^\circ$) and magnetic field azimuthal angle ($y-axis$, from $-180^\circ$ to $180^\circ$). The histograms are analogously structured and normalised as in Fig. \ref{fig:whistOverviewRight} except that here only the row for $0.95 \le\varpi_\mathrm{inj,P}\le 1.05$ is shown, the observations are restricted to the region of the focusing cone, and each column refers to observations associated with a different solar wind velocity $v_{\mathrm{SW}}$.}
\label{fig:whistFConeVsw}
\end{figure*}

\section{Definition of a injection-velocity based \pui velocity measure}
\label{ch:winjdef}
Table \ref{tab:wevo} provides a summary of the properties of \pui velocity measures and Fig. \ref{fig:vISNmodelshowcase}e displays $\varpi_\mathrm{SW}$ spectra as a function of ecliptic longitude. Similar to the findings of \cite{Moebius2015}, the velocity spectra are shifted as a function of ecliptic longitude. The reason for this systematic shift is that the velocity measure $\varpi_\mathrm{SW}$ of Eq.~\ref{eq:wrefswdef} does not account for the velocity of the \isn it originated from. The \isn velocity causes a shift of the initial velocity of \puis and due to orbital mechanics. Therein, the shift is a function of ecliptic longitude \cite[]{Moebius2015,Lee2015,Taut2018}. \cite{Moebius2015} corrected for the shift via a fit of a model curve to the cutoff of the orbital position-dependent $\varpi_\mathrm{SW}$ spectrum. This fit also yielded a then new way to determine the inflow direction of the local interstellar medium. \cite{Taut2018} increased the precision of the fit by taking into account various parameters including vector information of the orientation of $v_\mathrm{SW}$ and a model by \cite{Lee2015} of $v_\mathrm{ISN}$.

However, since in this study, the relative \pui velocity is computed from \threed vector information, the corrections proposed by \cite{Moebius2015,Taut2018} can not be employed directly, since they are ambiguous in direction. Therefore, a corrected $\varpi_\mathrm{SW}$-like quantity is required which becomes $1$ for freshly created \puis. Since the velocity distance of a \pui to the transporting \sw bulk $\vec{v}_\mathrm{PUI} - \vec{v}_\mathrm{SW}$ is initially $\vec{v}_\mathrm{SW} - \vec{v}_\mathrm{ISN}$, we propose, that the velocity measuring quantity
\begin{equation}
\varpi_\mathrm{SW,inj} = \frac{|\vec{v}_\mathrm{PUI} - \vec{v}_\mathrm{SW}|}{|\vec{v}_\mathrm{SW} - \vec{v}_\mathrm{ISN}|}
\end{equation}
suits the purpose of uniquely identifying fresh \puis with $\varpi_\mathrm{SW,inj}$ independent of ecliptic longitude. Since the computation of $\varpi_\mathrm{SW,inj}$ requires knowledge of $\vec{v}_\mathrm{ISN}$, computational means to compute the \isn trajectory and position-dependent velocity were implemented following \citet{Lee2015}, details are given in the Appendix \ref{ch:isnmod}. Table \ref{tab:wevo} compares this study's \pui velocity measure $\varpi_\mathrm{SW,inj}$ to the two most commonly used velocity measures for \puis.

Since for any point in the heliosphere two distinct trajectories lead to it, two different versions of $\varpi_\mathrm{SW,inj}$ are necessary ($\varpi_\mathrm{SW,inj,p}$ and $\varpi_\mathrm{SW,inj,s}$). Herein, p and s represent the primary and secondary trajectories, respectively. This distinction is determined based on the local \isn density under the assumption that the dominant process for reducing the \isn density is photoionization. The trajectory with the higher remaining \isn density at the considered orbital position is labelled as primary, the trajectory with the lower remaining \isn density as secondary. Fig. \ref{fig:vISNmodelshowcase}a shows the fraction of primary trajectory density to the total \isn density as a function of ecliptic longitude over the orbit of STEREO-A. Since the primary trajectory \isns make up more than $0.8$ of the total density everywhere except for $\sim \pm 30^\circ$ around the focusing cone (yellow triangle marker in Fig. \ref{fig:vISNmodelshowcase}), the secondary trajectory can be neglected outside the focusing cone region. Fig. \ref{fig:vISNmodelshowcase}b shows the radial and tangential velocity components of the \isn velocity. Close to the focusing cone, the primary and secondary become similar and, at the symmetry point, swap roles. Exactly at the focusing point, they are indistinguishable. 

In the following, we take $75^\circ$ as the \isn inflow direction which represents a compromise of the different values derived in \citet{Taut2018} and \citet{2022ApJ...937L..32S}.
Panels (c.) and (d.) of Fig.~\ref{fig:vISNmodelshowcase} display $\varpi_\mathrm{SW,inj}$ spectra as a function of ecliptic longitude analogously to the $\varpi_\mathrm{SW}$ spectra of Fig.~\ref{fig:vISNmodelshowcase}(e.). Here, $\varpi_\mathrm{SW,inj,p}$ and $\varpi_\mathrm{SW,inj,s}$ are computed under the assumption that all observed \puis followed the primary or the secondary component, respectively. Clearly, this assumption is unrealistic for the secondary trajectory outside the focusing cone region.  Thus, Fig.~\ref{fig:vISNmodelshowcase}(d.) shows a strong position-dependent velocity shift for the secondary trajectory outside the focusing cone region which is amplified compared to the standard definition of $\varpi_\mathrm{SW}$ in Panel (e.). Since outside the focusing cone region the contribution of particles following the secondary trajectory is expected to be negligible, almost only \puis of primary \isn origin are found. So here, the reference to the secondary injection point is not suitable. 
  
In contrast, the $\varpi_\mathrm{SW,inj,p}$ spectra in Fig.~\ref{fig:vISNmodelshowcase}(c.) align with their cutoff well to the horizontal line at $\varpi_\mathrm{SW,inj,p} = 1$. This illustrates  that $\varpi_\mathrm{SW,inj,p}$ indeed works as a relative velocity measure as intended. In particular,  $\varpi_\mathrm{SW,inj,p}$ now allows a direct comparison of \puis at different positions in STEREO-A's orbit.

\section{Traces of the torus of fresh \puis}
\label{sec:torus}

Based on 2d observations of STEREO-A/PLASTIC, \cite{Drews2015} found signatures that shifted in their observed direction as a function of the magnetic field direction. \cite{Drews2015} interpreted these observations as a torus of freshly-injected \puis (as illustrated here in Fig.~\ref{fig:PUIinjsketch}) and referred to this structure simply as a torus. We follow this nomenclature in this study and we want to emphasise that in this context torus only refers to the torus-signature of freshly-injected \puis. Here, we extend the analysis from \cite{Drews2015} based on 3d-observations and take our new injection-velocity dependent $\varpi_\mathrm{SW,inj}$ into account. 

For a direct comparison with \cite{Drews2015}, in this section we employ the same 2d angle $\beta$ (see Eq.~(3) in \cite{Drews2015}) of the relative \pui within the instrument's aperture. If $\varpi_\mathrm{SW}$ of Eq.~\ref{eq:wrefswdef} is generalised to a vector quantity $\vec{\varpi}_\mathrm{SW}$ by omission of the absolute in the numerator, then $\beta$ can be interpreted as the polar angle of $\vec{\varpi}_\mathrm{SW}$ in the $(x,y)$-plane, namely
\begin{equation}
\beta = \arctan\left(\frac{\varpi_\mathrm{SW,y}}{\varpi_\mathrm{SW,x}}\right)\ .
\label{eq:beta}
\end{equation}

Our current study incorporated the \isn velocity during pickup to increase the precision of the relative \pui velocity. Interestingly, modifications by the \isn pickup velocity do not affect $\beta$. This is evident from an analysis of the fraction of $\varpi_\mathrm{SW,inj,y} / \varpi_\mathrm{SW,inj,x}$ where the relation to the \isn velocity in the denominator cancels out like
\begin{align*}
  \frac{\varpi_\mathrm{SW,inj,y}}{\varpi_\mathrm{SW,inj,x}} =& \frac{\frac{v_\mathrm{PUI,y} - v_\mathrm{SW,y}}{|\vec{v}_\mathrm{SW} - \vec{v}_\mathrm{ISN}|}}{\frac{v_\mathrm{PUI,x} - v_\mathrm{SW,x}}{|\vec{v}_\mathrm{SW} - \vec{v}_\mathrm{ISN}|}}
  = \frac{v_\mathrm{PUI,y} - v_\mathrm{SW,y}}{v_\mathrm{PUI,x} - v_\mathrm{SW,x}}\\
  =& \frac{\frac{v_\mathrm{PUI,y} - v_\mathrm{SW,y}}{|\vec{v}_\mathrm{SW}|}}{\frac{v_\mathrm{PUI,x} - v_\mathrm{SW,x}}{|\vec{v}_\mathrm{SW}|}} = \frac{\varpi_\mathrm{SW,y}}{\varpi_\mathrm{SW,x}}\ .
\end{align*}
However, since \cite{Drews2015} employed a $\varpi_\mathrm{SW}$-filter to limit the data for their $(\beta,\varphi_\mathrm{B})$ histograms, while we here filter on $\varpi_\mathrm{SW,inj}$ our updated versions of these histograms still depend on the full 3d observations and are expected to differ from the \cite{Drews2015} version.

Fig.~\ref{fig:drewstorus} shows the angle $\beta$ over the magnetic field azimuthal angle $\vartheta_B$ for different filters in  $\varpi_\mathrm{SW,inj}$. We observe an analogous torus feature as seen in Fig.~5 in \citet{Drews2015}: The \pui observations peak at approximately $90^\circ$ and this peak shifts systematically with $\beta$ to higher azimuthal angles. We interpret this as a verification of the torus signature reported in \citet{Drews2015}.

\section{Pitch angle distributions as a function of magnetic field orientation and ecliptic longitude}
\label{ch:padresults}
This section presents the \ce{He+} \pui observations as pitch angle distributions as a function of the magnetic field azimuthal angle and sorted by the new \pui velocity measure. In the following, the magnetic field elevation angle is limited to in-ecliptic configurations of the magnetic field and only the $\varphi<0^\circ$ half of the SWS aperture is considered.

As an overview of the whole orbit of STEREO-A, Fig.~\ref{fig:whistOverviewRightGlobal} and Fig.~\ref{fig:whistOverviewRight} display pitch angle distributions as a function of magnetic field azimuthal angle and $\varpi_{\mathrm{SW,inj,P}}$, grouped by the ecliptic longitude they were observed at. The difference between Fig.~\ref{fig:whistOverviewRightGlobal} and Fig.~\ref{fig:whistOverviewRight} is the normalization of each histogram. In  Fig.~\ref{fig:whistOverviewRightGlobal} all histograms share the same global normalization, whereas in Fig.~\ref{fig:whistOverviewRight} all histograms are normalised to their respective 99-percentile. Fig.~\ref{fig:whistOverviewRightGlobal} clearly shows the expected enhancement of \puis near the focusing cone at $\lambda_{\mathrm{ecl}}=75^\circ$. The dashed lines in each histogram indicate the expected position of a torus of freshly injected \puis within PLASTIC's \fov and most histograms show a clear enhancement at this expected torus position. In addition, our newly proposed \pui velocity measure $\varpi_{\mathrm{SW,inj,P}}$ shows a strong population of \puis over all ecliptic longitudes in the range $0.85 \le \varpi_{\mathrm{SW,inj,P}} \le 0.95$. $\varpi_{\mathrm{SW,inj,P}}<0.95$ implies that these \puis might not be freshly injected \puis.

The individually normalised histograms in Fig.~\ref{fig:whistOverviewRight} allow a more detailed comparison of the shapes of the respective distributions. Common to all histograms in Fig.~\ref{fig:whistOverviewRight} are prominent features at pitch angles of $\sim 90^\circ$. These align with the regions where the range of expected pitch angles of freshly created \puis intersects the aperture (dashed lines). So, especially for $\varpi_{\mathrm{SW,inj,P}} \sim 1$, this can be interpreted as a strong signature of the torus of freshly created \puis. However, even though the features clearly have a maximum where it would be expected for freshly created \puis, the features are quite broad. Relative densities of $>0.25$ are observed for magnetic field azimuthal orientations in $75^\circ <|\varphi_B|<135^\circ$. Close to $\varphi_\mathrm{B} = 0^\circ$ and $\varphi_\mathrm{B} = \pm 180^\circ$ the relative density almost diminishes. This structure can at least partially be caused by the sensitivity range of the angular response function of the resistive anode of PLASTIC, and the blind spot of the instrument in the centre (compare Fig.~\ref{fig:ringcovdata}a). Hence, the possibility that additional anisotropic features are masked by the instrumental angular response and the pitch angle distributions, therefore, appear more isotropic can not be discounted. Yet, in the $\varpi_{\mathrm{SW,inj,P}} \sim 1$ row in Fig.~\ref{fig:whistOverviewRight}, it can be observed that the broadness of regions with relative densities $>0.75$ is increasing and decreasing with the ecliptic longitude.

 Next, the differences between the histograms observed at the same ecliptic longitude but different $\varpi_{\mathrm{SW,inj,P}}$ (different rows in Fig.~\ref{fig:whistOverviewRight}) are considered. Under the assumption that transport effects are reflected in a $\varpi_{\mathrm{SW,inj,P}}$ smaller or larger than $1$, then primarily fresh \puis are found at $\varpi_{\mathrm{SW,inj,P}} \sim 1$, and \puis which have undergone transport and are thus modified have values larger than 1 if they have been accelerated and smaller than 1 if have been decelerated \footnote{Consecutive acceleration and cooling processes that exactly cancel each other out, can also lead to $\varpi_{\mathrm{SW,inj,P}} \sim 1$.}. Under the assumption that only cooling and no acceleration occurred, the distance of $\varpi_{\mathrm{SW,inj,P}}$ to $1$ can be interpreted as an age of \puis with $\varpi_{\mathrm{SW,inj,P}} < 1$. For older \puis,  the magnetic field configuration at the point of ionization does not necessarily relate to the magnetic field configuration during observation of the particle. Hence, the range of initial pitch angles of \puis as a function of the magnetic field azimuthal angle as it is plotted in the histograms is probably only a correct indicator for $\varpi_{\mathrm{SW,inj,P}} \sim 1$ and is more difficult to interpret in the other rows of Fig.~\ref{fig:whistOverviewRightGlobal} and Fig.~\ref{fig:whistOverviewRight}.  Even without transport effects, the changing magnetic field conditions at each pick-up event would result in a broadening of the observed torus as a superposition of multiple tori oriented around different magnetic field directions.

 Despite these arguments, the bottom panel of Fig.~\ref{fig:whistOverviewRight} (slower \puis) looks interestingly similar to the central panel. On the one hand, this could imply that the time scale of \pui modification by transport is not sufficient to create significant differences between the panels and that $0.85<\varpi_{\mathrm{SW,inj,P}}<0.95$ is close enough to 1 for most \puis there to behave similarly to freshly injected \puis.
 On the other hand, also the \puis with  $\varpi_{\mathrm{SW,inj,P}} \sim 1$ might have already experienced transport effects that did not obscure the torus signature completely but already modified the pitch angle distribution in a similar manner as for the  \puis in $0.85<\varpi_{\mathrm{SW,inj,P}}<0.95$.  Further, \puis that were ionised close to but not directly at the spacecraft and under varying magnetic field conditions could also contribute to a broadening of the torus. However, we cannot rule out completely that the complex ageing effects and instrumental asymmetries together with insufficient statistics could obscure signatures of isotropic distributions \citep{Keilbach2023phd}.
  
The data in the top panels of Fig.~\ref{fig:whistOverviewRightGlobal} and Fig.~\ref{fig:whistOverviewRight} are more noisy, since fewer events contribute to the distributions. 
Despite this small statistic, it is interesting that the histograms of $\varpi_{\mathrm{inj},P} > 1$ show quite isotropic distributions, at least for all histograms outside the focusing cone. This tentatively hints at a possibly different transport behaviour of accelerated compared to cooled \puis. We expect that this can be observationally verified or falsified with future studies with the instrumentation on board Solar Orbiter, IMAP or Interstellar Probe that build on the methods employed in our current study. 

Fig.~\ref{fig:whistFconePS} displays pitch angle distributions filtered not only by $\varpi_\mathrm{SW,inj,P}$ which assumes that all \puis followed the primary trajectory, but for comparison, also for $\varpi_\mathrm{SW,inj,S}$ which assumes that all \puis followed the secondary trajectory.

At the focusing cone, based on Fig.~\ref{fig:vISNmodelshowcase} the velocities of the two \isn trajectories are expected to become similar and at the symmetry point swap their roles with regard to primary vs secondary trajectory. These expectations are tested here for $\varpi_\mathrm{SW,inj,P}$ and $\varpi_\mathrm{SW,inj,S}$ with regard to the similarity of the respective distributions at the focusing cone. Outside the focusing cone and crescent the probability of encountering particles from the secondary trajectory is very low due to the $r^{-2}$ dependency of ionization. The filters for $\varpi_{\mathrm{SW,inj,P}}$ or $\varpi_{\mathrm{SW,inj,S}}$ each treat all \puis as coming either from the primary or the secondary trajectory since of course the information on which of the two trajectories the \pui travelled prior to ionization is lost. Therefore, distinguishing between these two filters is only meaningful close to the focusing cone where both trajectories are relevant.  The insignificance of the secondary \isn trajectory outside the focusing cone is convenient, since again there is no way found to distinguish the origin of the \puis if the secondary trajectory were significant in density. Thus, at intermediate longitudes, it would be ambiguous but most important to decide whether to filter for $\varpi_{\mathrm{SW,inj,P}}$ or $\varpi_{\mathrm{SW,inj,S}}$, since there the trajectories' velocity vectors are most different from each other. Therefore, in all other figures in this study except Fig.~\ref{fig:whistFconePS}, the secondary trajectory is disregarded, since the corresponding histograms exhibit a misleading shift under the $\varpi_\mathrm{SW,inj,S}$ filter.

In Fig.~\ref{fig:whistFconePS}, each pair of the two pitch angle distributions within $10^\circ$ of the focusing cone's coordinate (centre of Fig.~\ref{fig:whistFconePS}) are almost identical at all displayed $\varpi_{\mathrm{SW,inj,P}}$ and $\varpi_{\mathrm{SW,inj,S}}$ intervals. With more distance from the focusing cone's centre, the $\varpi_{\mathrm{SW,inj,P}}$ and $\varpi_{\mathrm{SW,inj,S}}$ histograms become less similar to each other. At ecliptic longitudes not displayed, the difference becomes more distinct.  

That with the small differences between $\varpi_{\mathrm{SW,inj,P}}$ or $\varpi_{\mathrm{SW,inj,S}}$ in the focusing cone region still slightly different pitch angle distributions are found, emphasises the importance of a correct selection of the velocity of the transporting frame to relate the relative \pui velocity measure to.


Fig.~\ref{fig:whistFConeVsw} is restricted to the focusing cone and additionally filters by the solar wind speed $v_\mathrm{SW}$. For $250$ km/s $<v_\mathrm{SW}<$ $300$ km/s, the maximum for $|\varphi_\mathrm{B}|>0$ tends to be slightly closer to $0^\circ$ than the expected range of fresh \puis. This also applies to the range of $300$ km/s $<v_\mathrm{SW}<$ $350$ km/s. Then, with increasing $v_\mathrm{SW}$, a shift of the position of the maxima to higher pitch angles and $|\varphi_\mathrm{B}|$ is recognizable, which loosely correlates with $v_\mathrm{SW}$. In the last bin ($500$ km/s $<v_\mathrm{SW}<$ $550$ km/s) the statistical significance is low and the distribution looks more isotropic than in the other $v_\mathrm{SW}$ bins.

A definitive answer to whether there is a systematic relationship between the maxima in the pitch angle distributions and $v_\mathrm{SW}$ requires more data. Since our new \pui velocity measure $\varpi_{\mathrm{inj},P}$ makes \pui observation from the complete orbit of STEREO-A comparable, Fig.~\ref{fig:whistFullVsw} generalises Fig.~\ref{fig:whistFConeVsw} to the full orbit. In Fig.~\ref{fig:whistFullVsw}, which naturally includes more events than Fig.~\ref{fig:whistFConeVsw}, the positions of the maxima do not shift systematically with $v_\mathrm{SW}$. However, the distributions appear narrower for small values of $v_\mathrm{SW}$ and the widths of the distributions appear to increase with $v_\mathrm{SW}$. This small effect could be caused by different transport and acceleration conditions in different \sw regimes. For example, wave activity plays a stronger role for high solar wind velocities, whereas stream interaction regions which are acceleration sites for \puis are more frequent at intermediate solar wind velocities. Whether these processes indeed influence the \pui distributions is out of the scope of this paper and requires further investigation with future studies. Yet, since Fig.~\ref{fig:whistFConeVsw} mixes different orbital locations and multiple orbits (even different solar cycle states, since the orbits between 2008 and 2015 are observed) as well as different states of instrumental ageing, the broadening could also be a result of a mix of instrumental and physical effects. 


\section{Conclusions}
We reanalysed and recalibrated the STEREO-A PLASTIC \ce{He+} \pui data set. The recalibrated data set takes the full \threed \pui velocities into account and considers the unequal fraction of observable \puis depending on how much of a gyro-orbit falls into the \fov of PLASTIC. The resulting \pui distributions show a torus signature of freshly injected \puis and therein validates the results of \citet{Drews2015}.

We defined a new version of a \pui velocity measure, $\varpi_\mathrm{inj,}$ which corrects for the projection of the \isn velocity. This allows for the first time a direct comparison of \puis over the complete orbit of STEREO. The concept can also be applied to refine the determination of the \isn inflow direction from \citet{Moebius2015,Taut2018}.

We investigated pitch-angle distributions as a function of the magnetic field direction. Therein, the pitch-angle coverage is determined by PLASTIC's \fov and the magnetic field direction. This approach allows to observe unbiased partial pitch-angle distributions. 
The pitch-angle distributions are also sorted by our new \isn-velocity dependent $\varpi_\mathrm{inj,}$. Within the focusing cone we illustrated the symmetry between the primary and secondary trajectories that contribute to the \pui density.
 Over the complete orbit of STEREO, we observed torus-signatures of freshly-injected pick-up ions together with a probably already evolved but still similar population at lower $\varpi_\mathrm{inj,}$. Within the instrumental limitations (for example, ageing effects and potential contamination by protons, as discussed in detail in \citet{Keilbach2023phd}) of PLASTIC, these observations hint at the onset of transport effects. That a broadened torus signature is still visible for $0.85<\varpi_{\mathrm{SW,inj,P}}<0.95$  could possibly imply that these \puis were picked up upstream of the spacecraft and transport effects had not yet time to completely isotropise the distribution. For PLASTIC, the presented results are limited by the available statistics and can, at the moment, therefore not yet provide estimates of the time scales of scattering, cooling, and heating processes. Future observations with better statistics and, in particular, at different radial distances, for example with the Heavy Ion Sensor (HIS) as part of the Solar Wind Analyser (SWA, \citet{Owen2020}) on Solar Orbiter and the Solar Wind and Pick-up Ion (SWAPI) on the Interstellar Mapping and Acceleration Probe (IMAP, \citet{smith2024interstellar}), can benefit from our position independent \pui velocity measure and may provide the means to identify the radial position of the maximum of \ce{He} \puis which our results tentatively hint at being inside of 1 AU.



 \section*{Acknowledgements}
  This work was supported by the Deutsche Forschungsgemeinschaft (DFG) as WI 2139/12-1.  We further
  thank the science teams of STEREO/PLASTIC and STEREO/IMPACT for providing the respective level 1 and level 2 data
  products.

\bibliographystyle{aa}
\bibliography{bibliography.bib}

\begin{thebibliography}{34}
\expandafter\ifx\csname natexlab\endcsname\relax\def\natexlab#1{#1}\fi

\bibitem[{Acu{\~n}a {et~al.}(2008)Acu{\~n}a, Curtis, Scheifele, Russell,
  Schroeder, Szabo, \& Luhmann}]{Acuna2008}
Acu{\~n}a, M., Curtis, D., Scheifele, J., {et~al.} 2008, Space Science Reviews,
  136, 203

\bibitem[{{Axford}(1972)}]{Axford1972}
{Axford}, W.~I. 1972, {The Interaction of the Solar Wind With the Interstellar
  Medium}, ed. C.~P. {Sonett}, P.~J. {Coleman}, \& J.~M. {Wilcox}, Vol. 308,
  609

\bibitem[{{Biermann}(1957)}]{Biermann1957}
{Biermann}, L. 1957, Observatory, 77, 109

\bibitem[{{Drews} {et~al.}(2015){Drews}, {Berger}, {Taut}, {Peleikis}, \&
  {Wimmer-Schweingruber}}]{Drews2015}
{Drews}, C., {Berger}, L., {Taut}, A., {Peleikis}, T., \&
  {Wimmer-Schweingruber}, R.~F. 2015, \aap, 575, A97

\bibitem[{Drews {et~al.}(2012)Drews, Berger, Wimmer-Schweingruber, Bochsler,
  Galvin, Klecker, \& M{\"o}bius}]{drews2012inflow}
Drews, C., Berger, L., Wimmer-Schweingruber, R.~F., {et~al.} 2012, \jgr, 117

\bibitem[{{Galvin} {et~al.}(2008){Galvin}, {Kistler}, {Popecki}, {Farrugia},
  {Simunac}, {Ellis}, {M{\"o}bius}, {Lee}, {Boehm}, {Carroll}, {Crawshaw},
  {Conti}, {Demaine}, {Ellis}, {Gaidos}, {Googins}, {Granoff}, {Gustafson},
  {Heirtzler}, {King}, {Knauss}, {Levasseur}, {Longworth}, {Singer}, {Turco},
  {Vachon}, {Vosbury}, {Widholm}, {Blush}, {Karrer}, {Bochsler}, {Daoudi},
  {Etter}, {Fischer}, {Jost}, {Opitz}, {Sigrist}, {Wurz}, {Klecker}, {Ertl},
  {Seidenschwang}, {Wimmer-Schweingruber}, {Koeten}, {Thompson}, \&
  {Steinfeld}}]{Galvin2008}
{Galvin}, A.~B., {Kistler}, L.~M., {Popecki}, M.~A., {et~al.} 2008, \ssr, 136,
  437

\bibitem[{{Gloeckler} {et~al.}(2000){Gloeckler}, {Fisk}, {Geiss}, {Schwadron},
  \& {Zurbuchen}}]{Gloeckler2000}
{Gloeckler}, G., {Fisk}, L.~A., {Geiss}, J., {Schwadron}, N.~A., \&
  {Zurbuchen}, T.~H. 2000, \jgr, 105, 7459

\bibitem[{{Gloeckler} \& {Geiss}(1998)}]{Gloeckler1998}
{Gloeckler}, G. \& {Geiss}, J. 1998, \ssr, 86, 127

\bibitem[{{Gloeckler} {et~al.}(1993){Gloeckler}, {Geiss}, {Balsiger}, {Fisk},
  {Galvin}, {Ipavich}, {Ogilvie}, {von Steiger}, \& {Wilken}}]{Gloeckler1993}
{Gloeckler}, G., {Geiss}, J., {Balsiger}, H., {et~al.} 1993, Sci, 261, 70

\bibitem[{{Gloeckler} {et~al.}(2004){Gloeckler}, {M{\"o}bius}, {Geiss},
  {Bzowski}, {Chalov}, {Fahr}, {McMullin}, {Noda}, {Oka}, {Ruci{\'n}ski},
  {Skoug}, {Terasawa}, {von Steiger}, {Yamazaki}, \&
  {Zurbuchen}}]{Gloeckler2004}
{Gloeckler}, G., {M{\"o}bius}, E., {Geiss}, J., {et~al.} 2004, \aap, 426, 845

\bibitem[{{Gloeckler} {et~al.}(1995){Gloeckler}, {Schwadron}, {Fisk}, \&
  {Geiss}}]{Gloeckler1995h}
{Gloeckler}, G., {Schwadron}, N.~A., {Fisk}, L.~A., \& {Geiss}, J. 1995, \grl,
  22, 2665

\bibitem[{{Isenberg}(1997)}]{Isenberg1997}
{Isenberg}, P.~A. 1997, \jgr, 102, 4719

\bibitem[{{Jian} {et~al.}(2019){Jian}, {Luhmann}, {Russell}, \&
  {Galvin}}]{Jian2019}
{Jian}, L.~K., {Luhmann}, J.~G., {Russell}, C.~T., \& {Galvin}, A.~B. 2019,
  \solphys, 294, 31

\bibitem[{{Jian} {et~al.}(2018){Jian}, {Russell}, {Luhmann}, \&
  {Galvin}}]{Jian2018}
{Jian}, L.~K., {Russell}, C.~T., {Luhmann}, J.~G., \& {Galvin}, A.~B. 2018,
  \apj, 855, 114

\bibitem[{{Jian} {et~al.}(2013){Jian}, {Russell}, {Luhmann}, {Galvin}, \&
  {Simunac}}]{Jian2013}
{Jian}, L.~K., {Russell}, C.~T., {Luhmann}, J.~G., {Galvin}, A.~B., \&
  {Simunac}, K.~D.~C. 2013, in American Institute of Physics Conference Series,
  Vol. 1539, Solar Wind 13, ed. G.~P. {Zank}, J.~{Borovsky}, R.~{Bruno},
  J.~{Cirtain}, S.~{Cranmer}, H.~{Elliott}, J.~{Giacalone}, W.~{Gonzalez},
  G.~{Li}, E.~{Marsch}, E.~{Moebius}, N.~{Pogorelov}, J.~{Spann}, \&
  O.~{Verkhoglyadova}, 191--194

\bibitem[{Keilbach(2023)}]{Keilbach2023phd}
Keilbach, D. 2023, PhD thesis

\bibitem[{Lee {et~al.}(2012)Lee, Kucharek, M{\"o}bius, Wu, Bzowski, \&
  McComas}]{lee2012analytical}
Lee, M.~A., Kucharek, H., M{\"o}bius, E., {et~al.} 2012, ApJS, 198, 10

\bibitem[{{Lee} {et~al.}(2015){Lee}, {M{\"o}bius}, \& {Leonard}}]{Lee2015}
{Lee}, M.~A., {M{\"o}bius}, E., \& {Leonard}, T.~W. 2015, \apjs, 220, 23

\bibitem[{{Luhmann} {et~al.}(2008){Luhmann}, {Curtis}, {Schroeder}, {McCauley},
  {Lin}, {Larson}, {Bale}, {Sauvaud}, {Aoustin}, {Mewaldt}, {Cummings},
  {Stone}, {Davis}, {Cook}, {Kecman}, {Wiedenbeck}, {von Rosenvinge}, {Acuna},
  {Reichenthal}, {Shuman}, {Wortman}, {Reames}, {Mueller-Mellin}, {Kunow},
  {Mason}, {Walpole}, {Korth}, {Sanderson}, {Russell}, \&
  {Gosling}}]{Luhmann2008}
{Luhmann}, J.~G., {Curtis}, D.~W., {Schroeder}, P., {et~al.} 2008, \ssr, 136,
  117

\bibitem[{{McComas} {et~al.}(2009){McComas}, {Allegrini}, {Bochsler},
  {Bzowski}, {Christian}, {Crew}, {DeMajistre}, {Fahr}, {Fichtner}, {Frisch},
  {Funsten}, {Fuselier}, {Gloeckler}, {Gruntman}, {Heerikhuisen}, {Izmodenov},
  {Janzen}, {Knappenberger}, {Krimigis}, {Kucharek}, {Lee}, {Livadiotis},
  {Livi}, {MacDowall}, {Mitchell}, {M{\"o}bius}, {Moore}, {Pogorelov},
  {Reisenfeld}, {Roelof}, {Saul}, {Schwadron}, {Valek}, {Vanderspek}, {Wurz},
  \& {Zank}}]{McComas2009}
{McComas}, D.~J., {Allegrini}, F., {Bochsler}, P., {et~al.} 2009, Sci, 326, 959

\bibitem[{{McComas} {et~al.}(2004){McComas}, {Schwadron}, {Crary}, {Elliott},
  {Young}, {Gosling}, {Thomsen}, {Sittler}, {Berthelier}, {Szego}, \&
  {Coates}}]{McComas2004}
{McComas}, D.~J., {Schwadron}, N.~A., {Crary}, F.~J., {et~al.} 2004, \jgr, 109,
  A02104

\bibitem[{{M{\"o}bius} {et~al.}(1985){M{\"o}bius}, {Hovestadt}, {Klecker},
  {Scholer}, {Gloeckler}, \& {Ipavich}}]{Moebius1985}
{M{\"o}bius}, E., {Hovestadt}, D., {Klecker}, B., {et~al.} 1985, \nat, 318, 426

\bibitem[{{M{\"o}bius} {et~al.}(2015){M{\"o}bius}, {Lee}, \&
  {Drews}}]{Moebius2015}
{M{\"o}bius}, E., {Lee}, M.~A., \& {Drews}, C. 2015, \apj, 815, 20

\bibitem[{{Moebius} {et~al.}(1995){Moebius}, {Rucinski}, {Hovestadt}, \&
  {Klecker}}]{Moebius1995}
{Moebius}, E., {Rucinski}, D., {Hovestadt}, D., \& {Klecker}, B. 1995, \aap,
  304, 505

\bibitem[{{N{\v{e}}me{\v{c}}ek} {et~al.}(2020){N{\v{e}}me{\v{c}}ek},
  {{\v{D}}urovcov{\'a}}, {{\v{S}}afr{\'a}nkov{\'a}}, {N{\v{e}}mec}, {Matteini},
  {Stansby}, {Janitzek}, {Berger}, \& {Wimmer-Schweingruber}}]{Nemecek2020}
{N{\v{e}}me{\v{c}}ek}, Z., {{\v{D}}urovcov{\'a}}, T.,
  {{\v{S}}afr{\'a}nkov{\'a}}, J., {et~al.} 2020, \apj, 889, 163

\bibitem[{Owen {et~al.}(2020)Owen, Bruno, Livi, Louarn, Al~Janabi, Allegrini,
  Amoros, Baruah, Barthe, Berthomier, {et~al.}}]{Owen2020}
Owen, C., Bruno, R., Livi, S., {et~al.} 2020, \aap, 642, A16

\bibitem[{Shestakova(2015)}]{Shestakova2015}
Shestakova, L. 2015, Sol. Syst. Res., 49, 139

\bibitem[{Smith {et~al.}(2024)Smith, Kubota, Schwinger, \&
  Scherrer}]{smith2024interstellar}
Smith, E., Kubota, S., Schwinger, M., \& Scherrer, J. 2024, in 2024 IEEE
  Aerospace Conference, IEEE, 1--7

\bibitem[{{Sok{\'o}{\l}} {et~al.}(2016){Sok{\'o}{\l}}, {Bzowski}, {Kubiak}, \&
  {M{\"o}bius}}]{Sokol2016}
{Sok{\'o}{\l}}, J.~M., {Bzowski}, M., {Kubiak}, M.~A., \& {M{\"o}bius}, E.
  2016, \mnras, 458, 3691

\bibitem[{{Starkey} {et~al.}(2021){Starkey}, {Fuselier}, {Desai}, {Schwartz},
  {Russell}, {Wei}, {Madanian}, {Mukherjee}, \& {Wilson}}]{2021ApJ...913..112S}
{Starkey}, M.~J., {Fuselier}, S.~A., {Desai}, M.~I., {et~al.} 2021, \apj, 913,
  112

\bibitem[{{Swaczyna} {et~al.}(2022){Swaczyna}, {Schwadron}, {M{\"o}bius},
  {Bzowski}, {Frisch}, {Linsky}, {McComas}, {Rahmanifard}, {Redfield},
  {Winslow}, {Wood}, \& {Zank}}]{2022ApJ...937L..32S}
{Swaczyna}, P., {Schwadron}, N.~A., {M{\"o}bius}, E., {et~al.} 2022, \apjl,
  937, L32

\bibitem[{Tarnopolski \& Bzowski(2009)}]{Tarnopolski2009}
Tarnopolski, S. \& Bzowski, M. 2009, \aap, 493, 207

\bibitem[{{Taut} {et~al.}(2018){Taut}, {Berger}, {M{\"o}bius}, {Drews},
  {Heidrich-Meisner}, {Keilbach}, {Lee}, \& {Wimmer-Schweingruber}}]{Taut2018}
{Taut}, A., {Berger}, L., {M{\"o}bius}, E., {et~al.} 2018, \aap, 611, A61

\bibitem[{{Vasyliunas} \& {Siscoe}(1976)}]{VS1976}
{Vasyliunas}, V.~M. \& {Siscoe}, G.~L. 1976, \jgr, 81, 1247

\end{thebibliography}

\begin{appendix}
\section{Averaging and integration of STEREO-A IMPACT MAG Magnetic field data}
\label{ch:magavg}
To match the cadence of PLASTIC, the vector 8 Hz data is averaged for each 5 min measurement cycle of PLASTIC. This procedure synchronises the magnetic field measurements with the \pui \pha data. However, information is lost with regard to the detailed time evolution of the magnetic field and 5 min may not be an optimal time scale to obtain an average magnetic field relevant for the particles. At an absolute magnetic field of 5 nT, the gyration period of a \ce{He+} \pui is $\sim 52$ s, at 4 nT, it is $\sim 65$ s. Therefore, 1 min could be a more meaningful time period. One could also interpret the 5 min cadence of PLASTIC as an averaging over $\sim5$ gyro orbits. A detailed analysis of the impact of the averaging time period of the magnetic field is beyond the scope of this work. However, since the pitch angle is computed from the magnetic field orientation, a detailed look at the impact of the time window of averaging is an interesting topic for future studies and might increase the accuracy of this work's approach.

\begin{figure}
\includegraphics[width = \columnwidth]{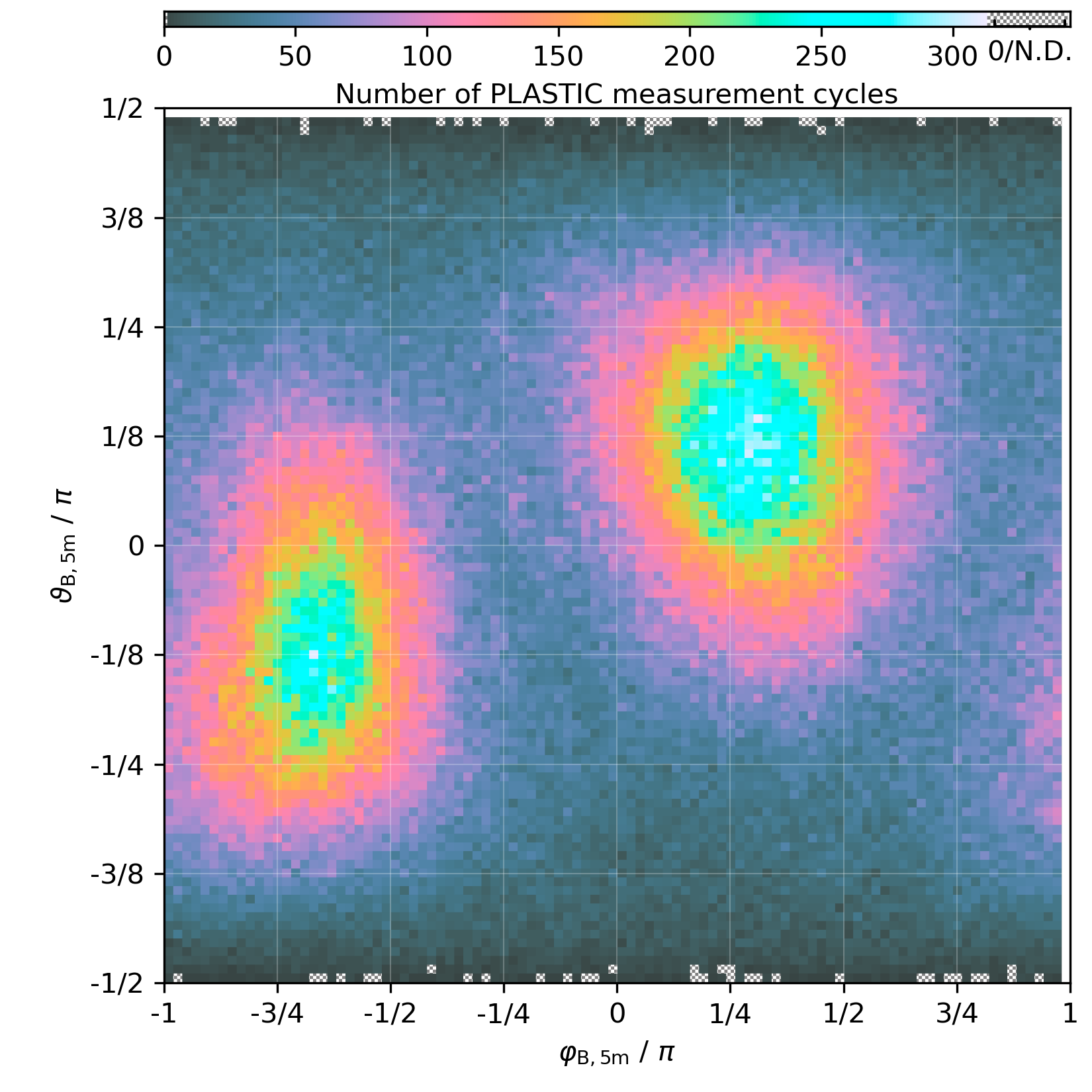}
\caption{Two dimensional histogram of the number of PLASTIC measurement cycles with magnetic field configurations selected by the azimuthal angle $\varphi_\mathrm{B}$ ($x$-axis) and the elevation angle $\vartheta_\mathrm{B}$ ($y$-axis).}
\label{fig:globalmagangles}
\end{figure}

The distribution of magnetic field angles resulting from 5 min averaged MAG data is shown in Fig. \ref{fig:globalmagangles}. This distribution is used as part of the normalization process depicted in Fig.~\ref{fig:ringcovdata}.

\section{Modelling of interstellar neutral trajectories}
\label{ch:isnmod}
A central ingredient for a precise assessment of the history of a \pui is knowledge of the velocity vector of the \isn it originated from. This section focuses on its derivation. 
\subsection{Derivation of interstellar neutral velocity vectors and densities}
\label{ch:isnvdef}

\begin{figure}
\includegraphics[width = \columnwidth]{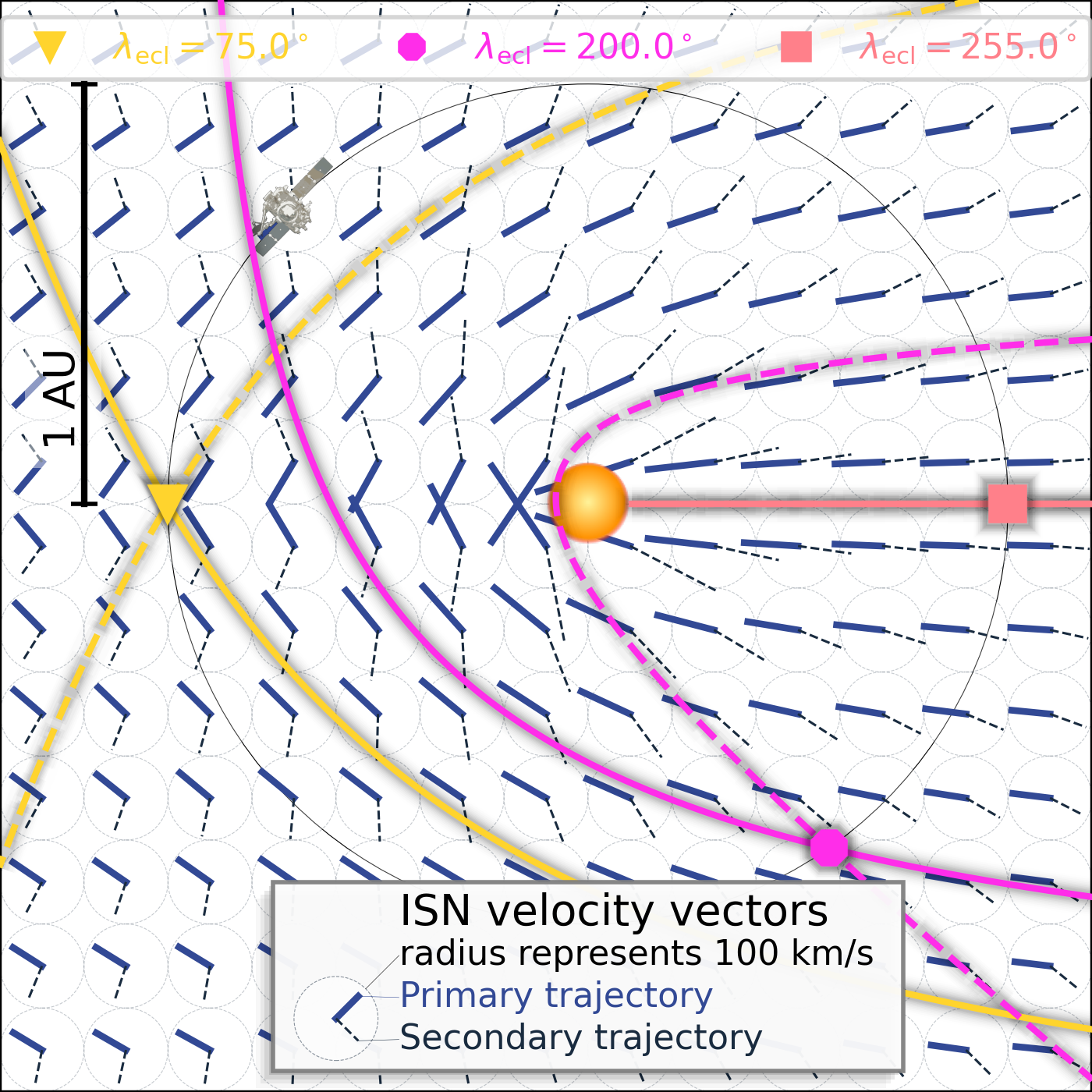}
\caption{Illustration of the \isn velocities for different positions.  The orbit of STEREO-A is indicated with a thin black line. Three pairs of example trajectories (primary with solid lines, and secondary with dashed lines) leading to the focusing cone (yellow), to the crescent (light pink), and an intermediate position (magenta) are shown. In addition, a grid of \isn velocities (in velocity space) is given for different positions in the inner heliosphere. The solid lines refer to the primary trajectory and the dashed lines to the secondary trajectory.}
\label{fig:vISNmodel}
\end{figure}

To obtain the velocity vectors of \isn neutrals along the orbit of STEREO-A at $\sim 1$ AU its direction at any point along its trajectory is required. The following derivation is similar to, for example, \citet{lee2012analytical}. The trajectory of a \ce{He} interstellar neutral is a Keplerian orbit (compare Sect.~\ref{ch:intro}) and obeys the vis-viva equation
\begin{equation}
v^2_\mathrm{ISN} = GM_\sun \left(\frac{2}{r} - \frac{1}{a}\right) \enspace.
\end{equation}
Here, $v_\mathrm{ISN}$ is the \isn velocity, $r$ the distance to the Sun, $a$ the semimajor axis of the trajectory, and $G$ the universal gravitational constant. For $r\rightarrow\infty$ and therefore $v_\mathrm{ISN} \rightarrow v_{\mathrm{ISN},\infty}$ the equation simplifies to
\begin{equation}
v^2_{\mathrm{ISN},\infty} = -\frac{GM_\sun}{a}, \text{which}
\end{equation}
for any $a$ does not necessarily have a real solution. However, if the condition $a < 0$ for hyperbolas is incorporated, $a$ may be replaced by $-|a|$, so that the negative sign vanishes on the right side. Then, 
\begin{equation}
|a| = \frac{\sqrt{GM_\sun}}{v_{\mathrm{ISN},\infty}}
\end{equation}
is obtained, so $a$ can be directly derived from the relative velocity between the heliosphere and the \isn, if the coordinate system is rotated such that the inflow direction coincides with the $x$-axis and the $xy$-plane is the plane in which the hyperbolic trajectory lies.

This is important for obtaining the semiminor axis $b$. While the absolute velocity at a solar distance of $r$ is already fully described by $a$ and $v_{\mathrm{ISN},\infty}$, the direction of the vector requires additionally the position at the orbit. For the hyperbolic trajectory in polar coordinates $(r,\varphi)$ the parametrization 
\begin{equation}
r(\varphi) = \frac{b^2/a}{1 - \varepsilon \cdot \cos(\varphi)}
\end{equation}
is employed. Herein, $\varepsilon = \sqrt{1 + \left(b/a\right)^2}$ is the eccentricity of the trajectory. The parametrization requires that the focus point of the hyperbola is the gravitational centre. Thus, in the following, the Sun is the origin of the coordinate system. Also, the parametrization requires that the periapsis coincides with the $-x$-direction. However, this is for most (but two) viewpoints at the orbit in contradiction to the earlier requirement that the $x$-direction is the direction of inflow. This can be solved by determining that the final trajectory is the result of a rotation around the $z$-axis of the parametrised trajectory. To determine the rotation angle, the initial angular coordinate of the parametrised trajectory is required. This angle, $\varphi_\infty$ is found through $r\rightarrow\infty$ and is computed by
\begin{equation}
\varphi_\infty = \arccos\left(-\frac{1}{\varepsilon}\right).
\end{equation}
Since the trajectory begins and ends at $\pm\varphi_\infty$, two possible trajectories are found per coordinate in the heliosphere. From geometrical deliberations, it is found that the rotation angle of the trajectory is $\varphi_R = \pm \varphi_\infty + \lambda_0$, where $\lambda_0$ is the inflow direction. Since $\varphi_\infty$ is determined by $\varepsilon$ and $\varepsilon$ is determined by $b$ and for $b$,  a point at the orbit in ecliptic longitude $\lambda$ and solar distance $r$ needs to be chosen and the problem becomes too complicated to be solved analytically. Instead, such a required point for the trajectory to coincide with is determined by the input coordinates $\lambda$ and $r$. The trajectory $r(\varphi)$ is dependent on $b$ through $\varepsilon$ and its rotation to match the required input coordinates is dependent on $b$ through $\varphi_\infty$. Hence, per orbit coordinate, $b$ is found numerically and as an optimization criterion, the minimum distance between the rotated trajectory and the required input point on the orbit is employed.

At this point, it is necessary to mention that by nature a Kepler trajectory is two dimensional. Here, it was implicitly assumed that the plane in which the trajectories are located is the in-ecliptic plane. As long as a cold model is assumed for the initial velocity of \isns, this assumption holds true for all trajectories except for the trajectories passing through in the focusing cone or crescent region. In a hot model (not employed here), the direction of inflow would vary slightly from particle to particle, since then the initial velocity is drawn from a non-delta distribution. The model can represent an out-of-ecliptic velocity component by allowing an additional rotation of the trajectory around the medium inflow axis. Similar deliberations lead to a model of focusing cone and crescent region inflow trajectories: At those symmetry points, the entirety of trajectories is found by a rotation of the in-ecliptic trajectories around the symmetry axis. In conclusion, this leads to the following statements about the \isn velocity vectors in a cold model:
\begin{itemize}
\item Per orbital coordinate $(r,\lambda)$ two trajectories are found. In the focusing cone and the crescent, the rotation of the two leads to the entirety of trajectories. Everywhere else, exactly two trajectories are found.
\item At a fixed distance from the Sun, the absolute \isn velocity is constant, however its direction is decisive for the initial \pui velocity.
\item For the \isn velocity vector in the sense of this work's model, only the radial (R) and tangential (T) components are relevant. Since in the focusing cone and the crescent region, the entirety of trajectories results from the rotation around the symmetry axis, the tangential and normal (N) components result from rotation of the original tangential component.
\end{itemize}
Since the \isn velocity components are readily obtained from the tangent of the computed trajectory, at this point two \isn velocity vectors per orbital coordinate are readily available. What remains, is a classification of the trajectories. For this, the probabilities of photoionization and charge exchange ionization of the \isns are numerically integrated. This results in density and ionization profiles along the trajectory. Therefore, the trajectory of higher density at the selected point along the spacecraft's orbit is classified as primary and the other as secondary. The resulting \isn velocities are illustrated in Fig.~\ref{fig:vISNmodel}. Three example trajectories are shown that pass through the focusing cone, the crescent, and an intermediate position in STEREO-A's orbit. The grid of velocity space representations show the local \isn velocities along the primary and secondary trajectories.

\section{Additional pitch-angle histograms}
\begin{figure*}[tp!]
\centering
\includegraphics[width = \textwidth]{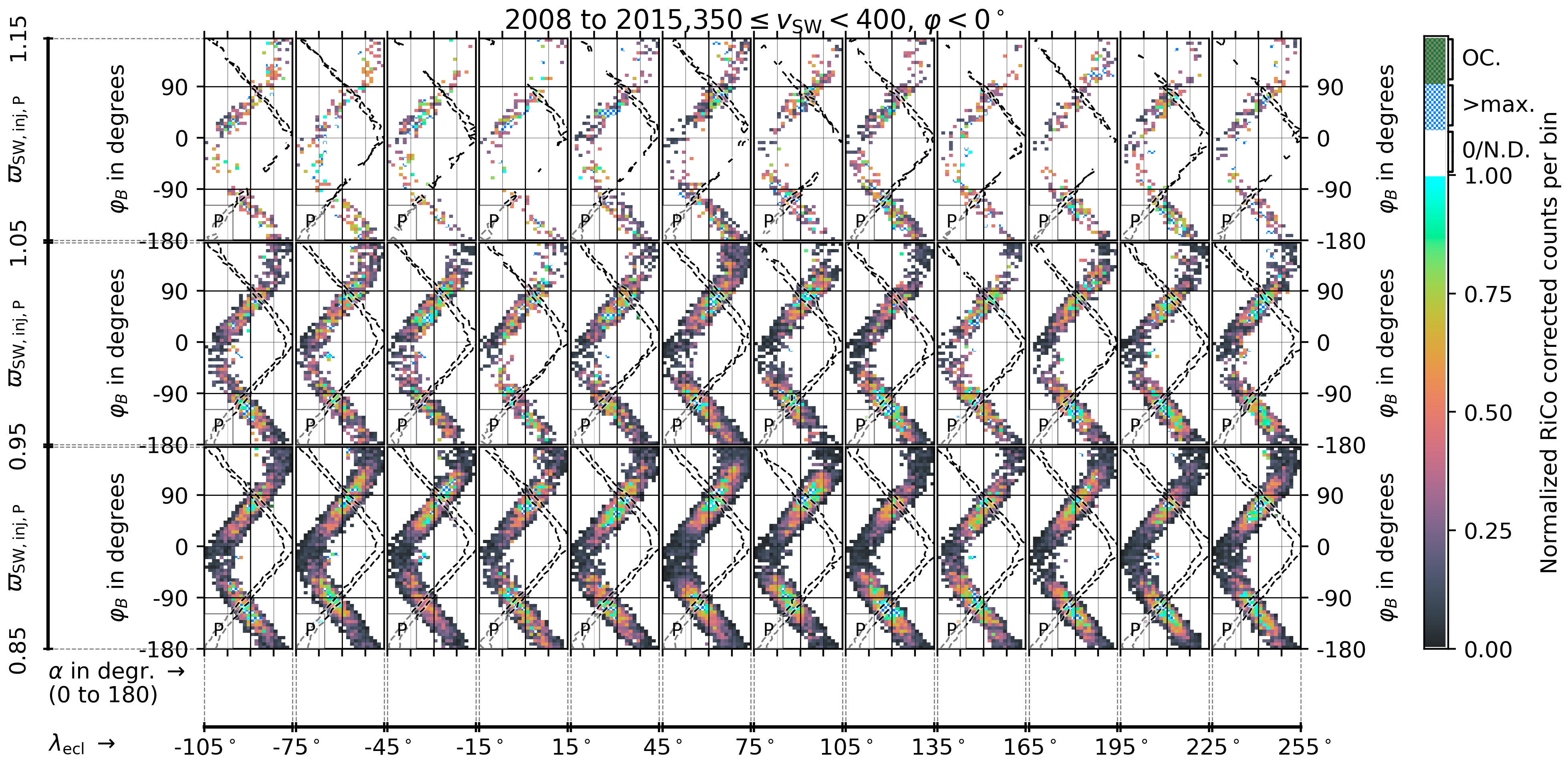}
\caption{2d histograms of pitch angle ($x-axis$, from $0^\circ$ to $180^\circ$ with one tick equalling $45^\circ$) and magnetic field azimuthal angle ($y-axis$, from $-180^\circ$ to $180^\circ$) in the same format as Fig~\ref{fig:whistOverviewRightGlobal} but here the histograms are normalised to their individual 99-percentile after the normalization steps displayed in Fig.~\ref{fig:ringcovdata} were applied.}
\label{fig:whistOverviewRight}
\end{figure*}

\begin{figure*}[tp!]
\centering
\includegraphics[width = \textwidth]{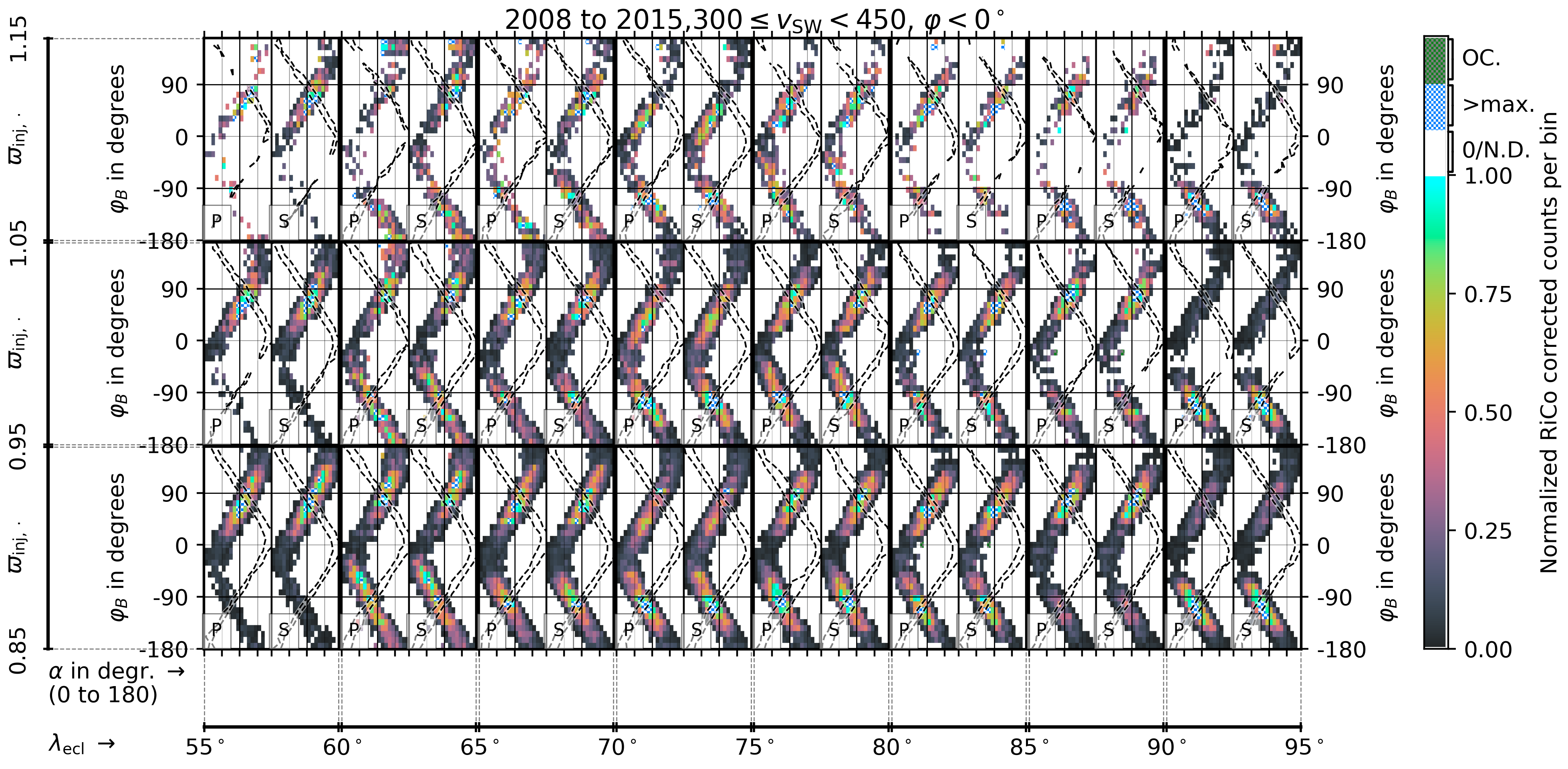}
\caption{2d histograms close to the focusing region of pitch angle ($x-axis$, from $0^\circ$ to $180^\circ$ with one tick equalling $45^\circ$) and magnetic field azimuthal angle ($y-axis$, from $-180^\circ$ to $180^\circ$) in the same format as Fig.~\ref{fig:whistOverviewRight} with the exception, that pairs of histograms individually filtered by the velocity of the primary (P) or the secondary (S) \isn trajectory are shown for each combination of $\varpi_\mathrm{inj,P}$/ $\varpi_\mathrm{inj,S}$ and ecliptic latitudes around the focusing cone.}
\label{fig:whistFconePS}
\end{figure*}

\begin{figure*}[tp!]
\centering
\includegraphics[width = \textwidth]{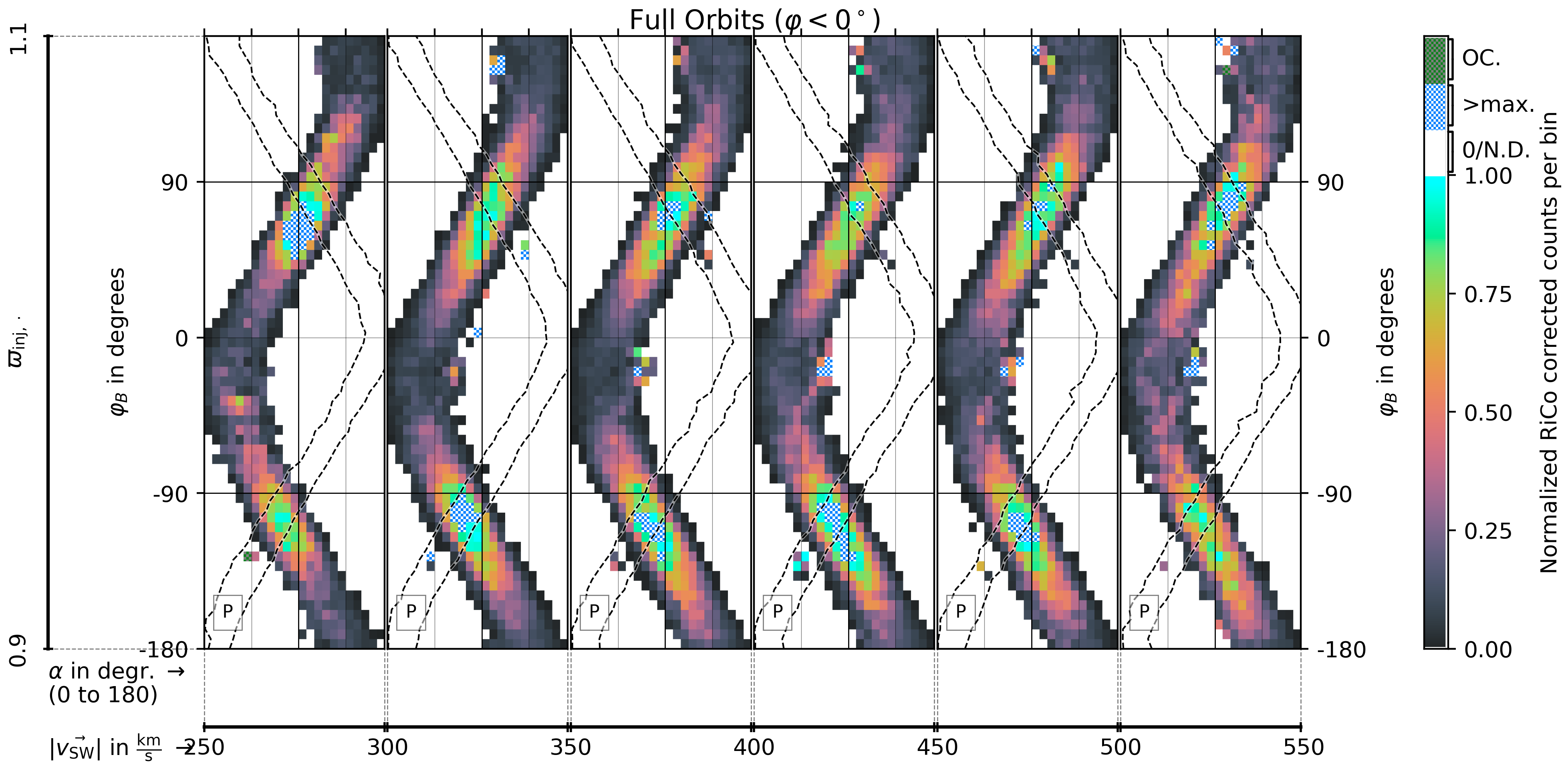}
\caption{2d histograms of pitch angle ($x-axis$, from $0^\circ$ to $180^\circ$ with one tick equalling $45^\circ$) and magnetic field azimuthal angle ($y-axis$, from $-180^\circ$ to $180^\circ$) in the same format as Fig.~\ref{fig:whistFConeVsw} but for full orbits.}
\label{fig:whistFullVsw}
\end{figure*}

This section supplies three additional figures that are part of the discussion in Sect.~\ref{ch:padresults}.

Fig.~\ref{fig:whistOverviewRight} show pitch-angle distribution over the orbit of STEREO-A in the same format as Fig.~\ref{fig:whistOverviewRightGlobal} but each subhistogram is normalised to its own maximum instead of the global maximum. 

Fig.~\ref{fig:whistFconePS} provides a detailed view of the focusing cone and compares the PUI velocity measure based on all particles following the primary trajectory $\varpi_{\mathrm{inj,P}}$ or the secondary trajectory $\varpi_{\mathrm{inj,S}}$.

Finally, Fig.\ref{fig:whistFullVsw} shows the pitch-angle distributions organised by the solar wind speed over the full orbit of STEREO-A.
\end{appendix}
\end{document}